\documentclass[twocolumn,aps,showpacs]{revtex4-1}
\usepackage[latin9]{inputenc}
\usepackage{amsmath}
\usepackage{amssymb}
\usepackage{graphicx}

\makeatletter

\providecommand{\tabularnewline}{\\}

\usepackage{subcaption}
\usepackage{color}

\def\NOT(#1,#2){\OneQubitGate(#1,#2){$X$}}

\makeatother
\begin{document}

\title{Efficient Implementation of a Quantum Algorithm in a Single Nitrogen
Vacancy Center of Diamond}

\author{Jingfu Zhang, Swathi S. Hegde and Dieter Suter\\
 Fakultaet Physik, Technische Universitaet Dortmund,\\
 D-44221 Dortmund, Germany}

\date{\today}

\begin{abstract}
Quantum computers have the potential to speed up certain problems
that are hard for classical computers. Hybrid systems, such as the
nitrogen vacancy (NV) center in diamond, are among the most promising
systems to implement quantum computing, provided the control of the
different types of qubits can be efficiently implemented. In the case
of the NV center, the anisotropic hyperfine interaction allows one
to control the nuclear spins indirectly, through gate operations targeting
the electron spin, combined with free precession. Here we demonstrate
that this approach allows one to implement a full quantum algorithm,
using the example of Grover's quantum search in a single NV center,
whose electron is coupled to a carbon nuclear spin. 
\end{abstract}
\maketitle

\textit{Introduction.-} Storing and processing digital information
in quantum mechanical systems has an enormous potential for solving
certain computational problems that are intractable in classical computers
\cite{nielsen,Stolze:2008xy}. Important examples of efficient algorithms
that require quantum mechanical processors include Grover's quantum
search \cite{PhysRevLett.79.325} over an unsorted database and prime
factorization using Shor's algorithm \cite{doi:10.1137/S0097539795293172}.
Hybrid systems consisting of different types of physical qubits, such
as the nitrogen vacancy (NV) center in diamond, appear promising for
building quantum computers \cite{ladd2010quantum,blencowe2010quantum,cai2014hybrid,kurizki2015quantum,Suter201750},
since they combine useful properties of different types of qubits.
The NV center \cite{Wrachtrup_2006,Doherty:2013uq,Suter201750}, e.g.,
combines the long coherence time of the nuclear spins with the rapid
operations possible on the electron spins. However, the benefits are
limited by the fact that the coupling between the nuclear spins and
the external control fields is 3-4 orders of magnitudes weaker than
for the electron spins, which results in slow operations of the nuclear
spins if the gates are implemented by control fields based on radio-frequency
(RF) pulses \cite{dobrovitski2012,PhysRevLett.115.110502}.

The strategy of indirect control \cite{PhysRevA.78.010303,PhysRevA.76.032326,PhysRevLett.107.170503,PhysRevLett.102.210502,PhysRevA.91.042340,PhysRevLett.109.137602,naturephoton,PhysRevA.96.032314,PhysRevB.96.134314,Pan13,zhang18,swathi19}
can reduce this limitation. This approach does not require external
control fields (RF pulses) acting directly on the nuclear spins. Instead,
only microwave (MW) pulses acting on the electron spin are applied,
combined with free precession under the effect of anisotropic hyperfine
interactions between the electron and nuclear spins. In previous works,
we used this approach for the implementation of basic operations like
initialization of qubits and quantum gate operations, including a
universal set of gates for quantum computing \cite{zhang18,swathi19}.
In these works, we could greatly improve the control efficiency, e.g.,
compared with approaches based on multiple dynamical decoupling cycles
\cite{PhysRevLett.109.137602,naturephoton,PhysRevB.96.134314} or
modulated pulses \cite{PhysRevA.78.010303,PhysRevLett.107.170503}:
our elementary unitary operations consisted of only 2 - 3 rectangular
MW pulses separated by delays.

Here, we apply this approach to the implementation of a full quantum
algorithm, Grover's search algorithm\cite{PhysRevLett.79.325}, which
is one of the milestones in the field of quantum information. In the
task of finding one entry in an unsorted database, Grover's search
algorithm scales with the size $N$ of the database as ${\cal O}\left(\sqrt{N}\right)$,
while all classical algorithms scale as ${\cal O}\left(N\right)$.
Grover's quantum search has been implemented in various physical systems,
such as NMR \cite{PhysRevLett.80.3408,ChaungGrover3,zhang:042314},
NV centers \cite{dobrovitski2012,groverdu}, trapped atomic ions \cite{PhysRevA.72.050306,grover3nc},
optics \cite{PhysRevLett.88.137901} and superconducting systems \cite{groversup}.
In this work, we implement it by indirect control, with only 4 MW
pulses for the whole quantum search. The experimental results demonstrate
the very high efficiency of the indirect control in implementing quantum
computing.

\textit{Grover's quantum search.-} Grover's search algorithm \cite{PhysRevLett.79.325}
can speed up the search of an unsorted database quadratically compared
to the classical search. The algorithm starts by initializing the
$n$- qubit quantum register to an equal superposition of all basis
states, 
\[
|\Psi\rangle_{in}=\frac{1}{\sqrt{N}}\sum_{i=0}^{N-1}|i\rangle,
\]
where $N=2^{n}$ and $|i\rangle$ denote the basis states of the system,
each of which maps to an item in the database. This state can be prepared
by initializing all qubits into state $|0\rangle$ and then applying
Hadamard gates ($H^{\otimes n}$) to each of them.

The algorithm then requires the repeated application of two operations
$D$ and $I_{t}$, where the oracle $I_{t}$ implements a phase flip
operation for the target state $|s_{t}\rangle$ but does not change
any other state: $I_{t}=I-2|s_{t}\rangle\langle s_{t}|$, where $I$
is the identity operator in the $n$- qubit system. $D$ denotes a
diffusion operation, and can be represented as $D=2P-I=H^{\otimes n}I_{|00\dots0\rangle}H^{\otimes n}$,
where$|00\dots0\rangle$ denotes the state of all qubits in $|0\rangle$
and $P=(\sum_{i,j=0}^{N-1}|i\rangle\langle j|)/N$. After applying
$U=DI_{t}$ to $|\Psi\rangle_{in}$ $m$ times, the system is in the
state $|\Psi\rangle_{out}=U^{m}|\Psi\rangle_{in}$. In this state,
the amplitude of the target state can approach 1 after $m\sim{\cal O}\left(\sqrt{N}\right)$,
while a classical search requires ${\cal O}\left(N\right)$ oracle
operations.

\begin{figure}[h]
\centering{}\includegraphics[width=1\columnwidth]{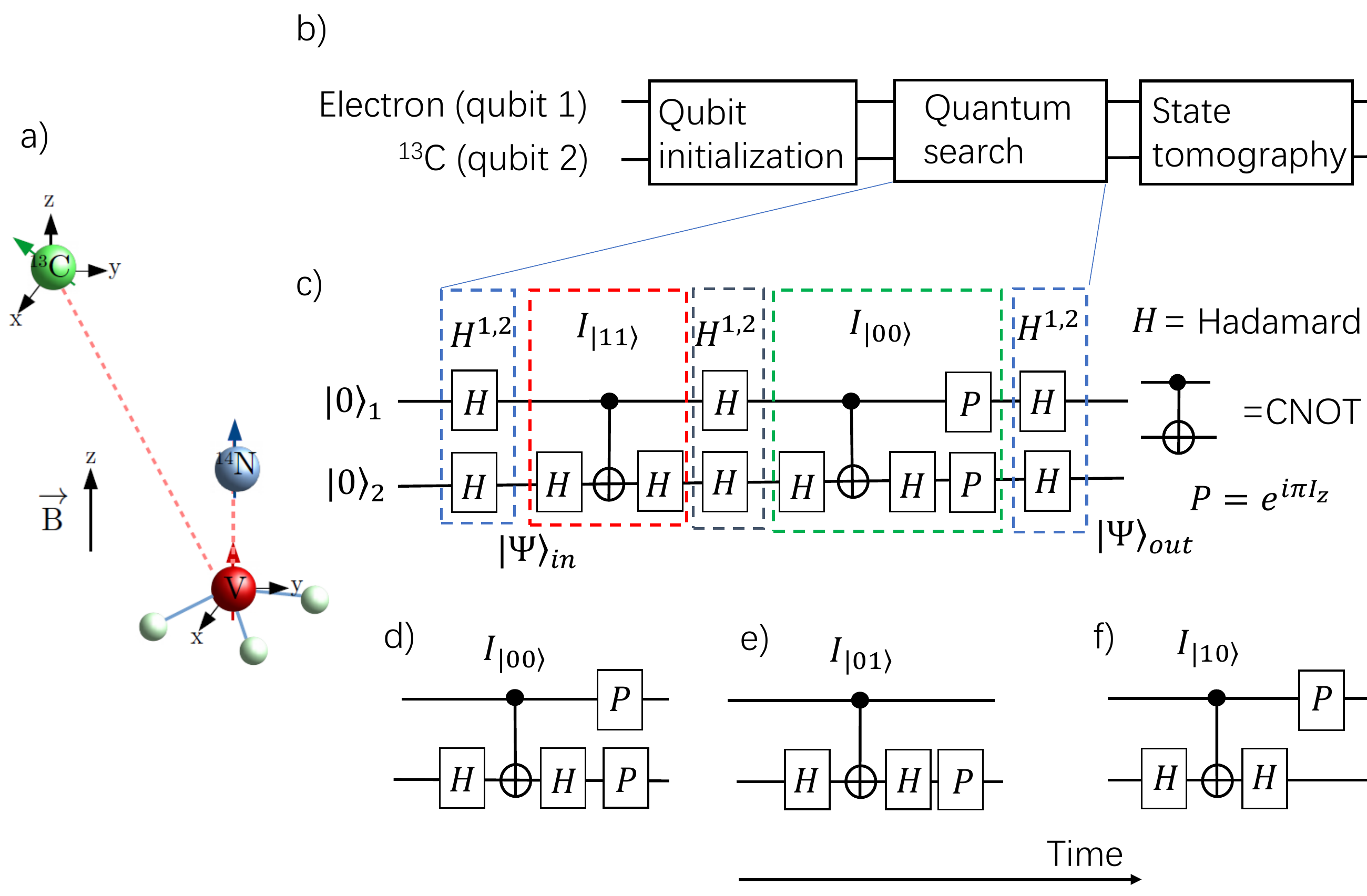} \caption{ (a) Structure of the NV system with the electron spin coupled to
one $^{14}$N and one $^{13}$C nuclear spin. (b) Schematic representation
of the experimental procedure, including the initialization, the quantum
search and the state tomography for determining the outcome, where
(c) shows the gate sequence for searching the target state $|11\rangle$.
$|\Psi\rangle_{in}$ denotes the equal superposition of all basis
states. The output state $|\Psi\rangle_{out}$ is the target state
$|11\rangle$. The circuits for the other target states $|00\rangle$,
$|01\rangle$, $|10\rangle$ are obtained by replacing the phase flip
operation $I_{|11\rangle}$ by $I_{|00\rangle}$, $I_{|01\rangle}$,
$I_{|10\rangle}$, which can be implemented by the circuits shown
in (d-f). \label{figoutline}}
\end{figure}

\textit{Experimental protocol.-} For the experimental implementation
we used a diamond with 99.995\% $^{12}$C, and the concentration of
substitutional nitrogen of $<10$ ppb to minimize decoherence \cite{doi:10.1063/1.4731778,Teraji_2013,PhysRevLett.110.240501}.
The experimental setup is presented in the SM. The experiment was
performed at room temperature in a static magnetic field $B$ of 14.8
mT along the symmetry axis of the NV center. The structure of the
NV center with the coupled $^{14}$N and $^{13}$C nuclear spins is
illustrated in Figure \ref{figoutline} (a). Here we use a symmetry-adapted
coordinate system, where the $z$-axis is oriented along the NV axis,
while the $^{13}$C nucleus is located in the $xz$-plane \cite{PhysRevB.94.060101}.
In this context, we focus on the subsystem where the $^{14}$N is
in the state $m_{N}$=1. The relevant Hamiltonian for the electron
and $^{13}$C spins is then 
\begin{equation}
\frac{\mathcal{H}_{e,C}}{2\pi}=DS_{z}^{2}-(\gamma_{e}B-A_{N})S_{z}-\gamma_{C}BI_{z}+A_{zz}S_{z}I_{z}+A_{zx}S_{z}I_{x}.\label{eq:Ham2spin}
\end{equation}
Here $S_{z}$ denotes the spin-1 operator for the electron and $I_{x/z}$
the $^{13}$C spin-1/2 operators. The zero-field splitting is $D=2.87$
GHz. $\gamma_{e/C}$ denote the gyromagnetic ratios for the electron
and $^{13}$C spins, respectively. $A_{N}=-2.16$ MHz is the secular
part of the hyperfine coupling between the electron and the $^{14}$N
nuclear spin~\cite{PhysRevB.89.205202,PhysRevB.47.8816,Yavkin16},
while $A_{zz}$ and $A_{zx}$ are the relevant components of the $^{13}$C
hyperfine tensor, which are $A_{zz}=-0.152$ MHz and $A_{zx}=0.110$
MHz in the present system.

We select a 2 qubit system for implementing the quantum search by
focusing on the subspace with the electron spin in $\{|0\rangle,|-1\rangle\}$
as qubit 1 and the $^{13}$C spin as the second qubit. Our computational
basis $\{|0\rangle,|1\rangle\}_{1}\otimes\{|0\rangle,|1\rangle\}_{2}$
corresponds to the physical states $\{|0\rangle,|-1\rangle\}_{e}\otimes\{|\uparrow\rangle,|\downarrow\rangle\}_{C}$,
where the states $|0\rangle$ and $|-1\rangle$ denote the eigenstates
of $S_{z}$, $|\uparrow\rangle$ and $|\downarrow\rangle$ the eigenstates
of $I_{z}$ with eigenvalues of $1/2$ and $-1/2$, respectively.
Figure \ref{figoutline} (b) outlines the protocol for implementing
the quantum search.

In the step of qubit initialization, we use a 4 $\mu s$, 0.5 mW pulse
of 532 nm laser light to initialize the electron spin into the $m_{S}=0$
state. Additional details of the setup are presented in the Supplemental
Material (SM) \cite{note_SM}, which includes Refs. \cite{Childress281,Ournewpaper,RevModPhys.76.1037,PhysRevA.98.052354}. Based on the initialized electron spin, we further
polarize the $^{13}$C spin by a combination of MW and laser pulses,
and set the qubits into the pure state $|00\rangle$ \cite{zhang18,swathi19}.
Additional details are given in the SM.

The protocol for the actual quantum search is shown in Figure \ref{figoutline}
(c) for the target state $|s_{t}\rangle=|11\rangle$. The circuits
for the other target states $|00\rangle$, $|01\rangle$, and $|10\rangle$
are obtained by replacing the phase flip operation $I_{|11\rangle}$
by $I_{|00\rangle}$, $I_{|01\rangle}$, and $I_{|10\rangle}$, as
shown in Figure \ref{figoutline} (d-f), respectively.

To implement the actual search shown as Figure \ref{figoutline} (c),
we considered sequences of MW pulses with constant MW and Rabi frequencies
but variable durations and phases. The MW frequencies were resonant
with the ESR transitions between the electron states $m_{S}=0\leftrightarrow m_{S}=-1$.
The pulse durations, phases and delays were used as variables in an
optimization procedure based on optimal control (OC) theory \cite{Mitchell:1998:IGA:522098,zhang18,swathi19}
that maximizes the overlap between the operation generated by the
sequence and the operation required by the quantum circuit of Figure
\ref{figoutline} (c), which can reach unity in the ideal case \cite{bookcontrol,PhysRevA.78.010303,PhysRevA.76.032326}.

The OC process has to balance several considerations. While it is
helpful to use many pulses and therefore many degrees of freedom to
optimize the theoretical fidelity of the gate operaation, additional
pulses also increase the total duration of the sequence and therefore
the effect of decoherence (mainly from the electron spin in the present
work), and the experimental imperfections also tend to increase with
the number of pulses. We found that sequences of 4 pulses and 5 delays
to be a good compromise for all four target states, see details in
SM.

To determine the state of the system after the search operation, we
use the techniques developed in quantum state tomography \cite{nielsen,PhysRevA.69.052302},
to reconstruct the populations or full quantum states. This requires
a set of measurements applied to the output state $|\Psi\rangle_{out}$.

\begin{figure}
\centering{}\includegraphics[width=1\columnwidth]{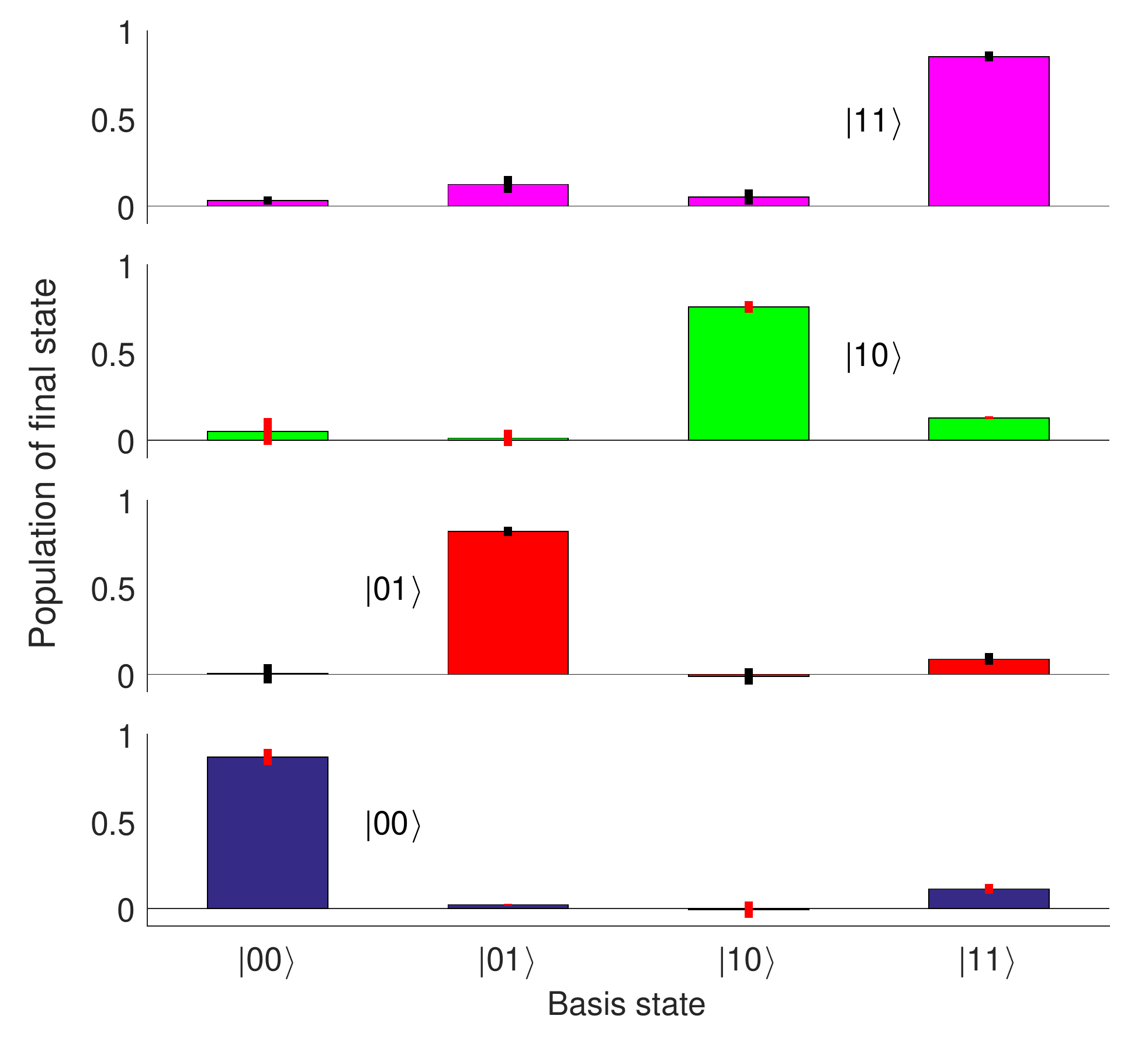} \caption{Experimentally measured populations of the four basis states after
the quantum search with the targets $|00\rangle$, $|01\rangle$,
$|10\rangle$, and $|11\rangle$, as indicated in the panels. The
error bars indicate 1 standard deviation, which was determined by
repeating each experiment.\label{figpopu}}
\end{figure}

\textit{Experimental results.-} Figure \ref{figpopu} illustrates
the experimental results for the different target states. Here we
only show the populations obtained from partial tomography. The measured
populations of the target states, or the probabilities of finding
the target states, are $0.87\pm0.05$, $0.82\pm0.03$, $0.76\pm0.03$
and $0.85\pm0.03$ for the target states $|00\rangle$, $|01\rangle$,
$|10\rangle$ and $|11\rangle$, with the sums of the populations
$1.00\pm0.03$, $0.90\pm0.03$, $0.95\pm0.06$ and $1.06\pm0.02$,
respectively. In each case the probability of finding the target state
is much higher than the classical result of $0.25$. In the ideal
case, the population of the target state should be 1 and the others
0. The deviation of the sum of the populations for each case from
the unity can be mainly attributed to imperfections in the tomography,
which cause population leakage to the electron state $m_{S}=1$. Secondly,
the incomplete selectivity of the MW pulses leads to loss of population
from the computational subspace. We estimate that this contribution
is less than 0.027 in our experiments. The effects from the coupled
$^{14}$N spin can be decreased, e.g., by polarizing $^{14}$N \cite{PhysRevApplied.12.024055,NJP19073030,PhysRevA.101.013835,Appl.Phys.Lett.105.242402,nature_514_72,PolN14Duan2,Yun_2019},
where the polarization can be $>98\%$ \cite{PhysRevApplied.12.024055,Yun_2019}.
The details are presented in the SM.

\begin{figure}
\centering{}\includegraphics[width=1\columnwidth]{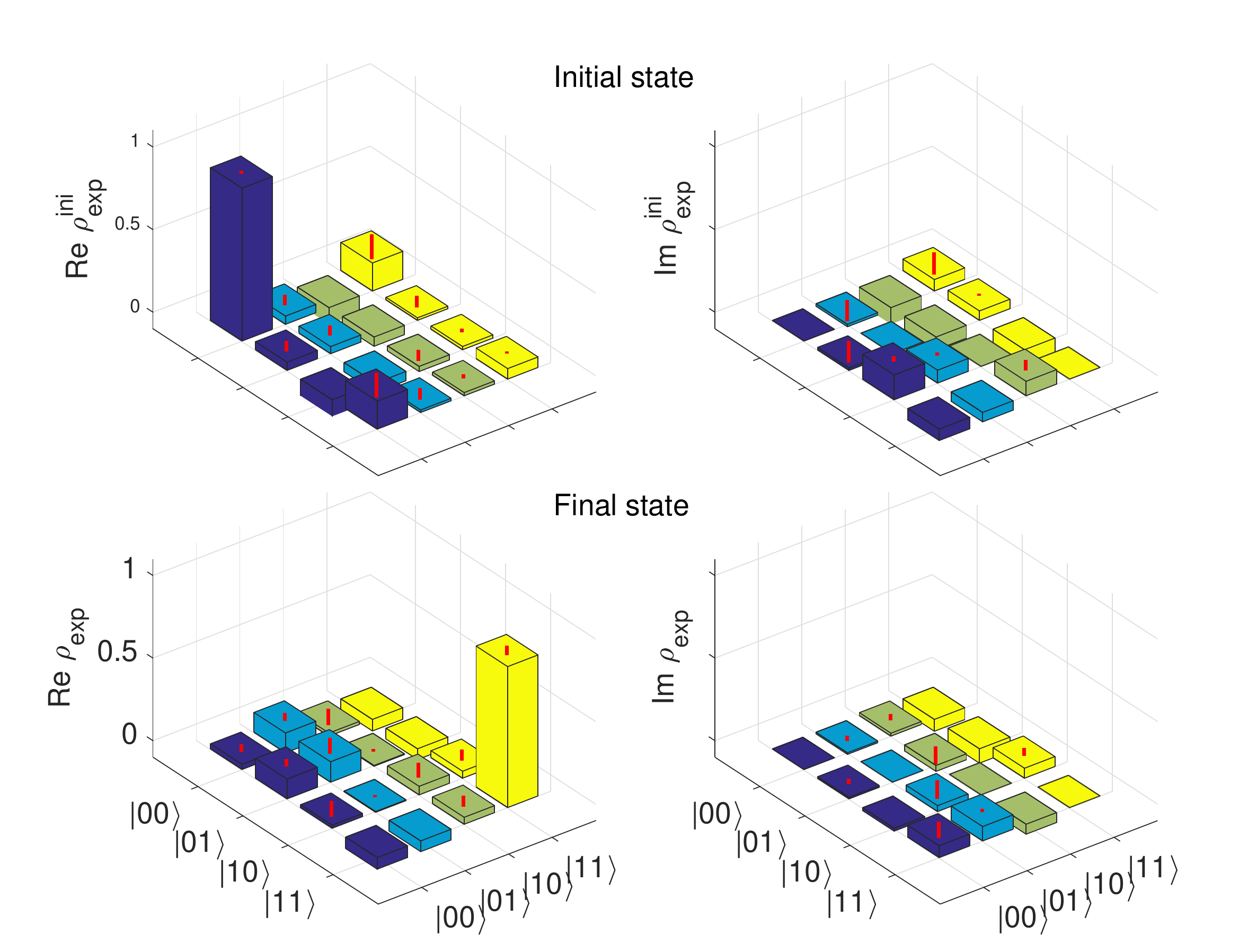} \caption{Measured density matrices for the prepared initial state $|00\rangle$
(top) and for the state after the completion of the quantum search
with target state $|11\rangle$ (bottom). The real and imaginary parts
are shown in the left and right columns. \label{figfull}}
\end{figure}

Figure \ref{figfull} shows the reconstruction of the full density
matrices for the initial state and the search result for the target
state $|11\rangle$. By calculating the fidelities as $F=Tr\{\rho_{th}\rho_{exp}\}$,
we obtained $F_{ini}=0.92\pm0.01$, and $F_{|11\rangle}=0.85\pm0.03$,
for the initial state and the final state after the quantum search.
The loss of the fidelity in the quantum search can be attributed to
the imperfection of the theoretical pulse sequence, the experimentally
implemented sequence and the experimentally implemented initial state
including the state tomography, in the order of importance. The details
of the error estimation are presented in the SM.

\textit{Discussion.-} The OC efficiency can be improved by maximizing
the angle between the different quantization axes of the nuclear spin
for the different states of the electron spin \cite{PhysRevA.76.032326}.
The experimental fidelity might be improved further, e.g., by increasing
the robustness with respect to fluctuations of the Rabi frequency
(see examples in the SM, Section VIIA), and increasing the Rabi frequency
\cite{zhang18}, e.g., in the case that the $^{14}$N is polarized.
Moreover, the choice of a more efficient optimal algorithm should
be helpful \cite{arxiv2003.03776,Weise09}. To estimate the scalability
of the OC scheme in larger systems, we use numerical simulation of
systems with one electron spin and $n=1$, ..., 4 $^{13}$C spins.
As examples, we use 3-4 MW pulses with 4-5 delays to implement the
CNOT-like gates, where the electron spin (in the subspace $m_{S}=0$
and $m_{S}=-1$) acts as the control qubit, while one $^{13}$C spin
is the target qubit. The target operation is chosen as $R_{x}^{j}(\pi)=e^{-i\pi I_{x}^{j}}$,
where $j$ indicates the target $^{13}$C spin. The details are presented
in the SM.

\begin{figure}
\centering{}\includegraphics[width=1\columnwidth]{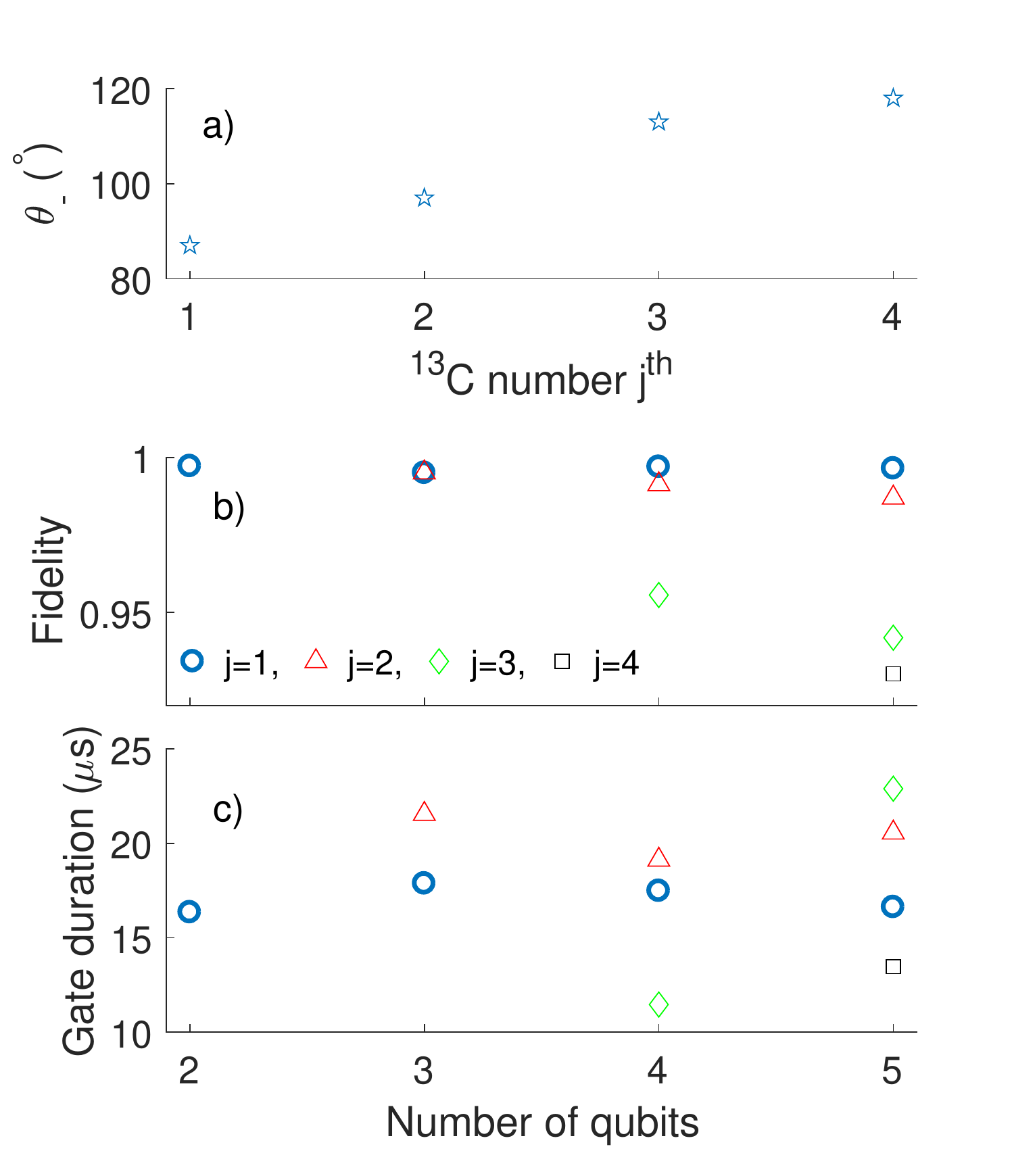} \caption{(a) The quantization axis orientation of $j$th $^{13}$C spin of
the subspace $m_{S}=-1$ between $z$- axis, the orientation of the
subspace $m_{S}=0$. (b-c) Results by simulation of controlled- $R_{x}^{j}(\pi)$
(CNOT-like gates) in 2-5 qubit systems, respectively, where the electron
spin is the control qubit, and $R_{x}^{j}(\pi)=e^{-i\pi I_{x}^{j}}$
with $j$ indicating the affected $^{13}$C spin. Figures (b-c) show
the gate fidelity and duration, respectively. \label{figmultC}}
\end{figure}

We investigate the dependence of the gate fidelity and duration on
the number of the qubits in the system. The results are shown in Figure
\ref{figmultC}, and the parameters for the pulse sequences are presented
in the SM. The results show that the $^{13}$C spin quanzitation axis
orientation in the subspace $m_{S}=-1$, denoted as $\theta_{-}$
in Figure \ref{figmultC} (a), is a crucial factor in the optimization.
The quality of the gate, here evaluated by the gate fidelity and duration,
is degraded only marginally by the passive $^{13}$C spins coupled
to the electron. For example, for $j=1$, with $\theta_{-}=87{}^{\circ}$,
the gate fidelity is higher than $0.995$, and the gate duration remaims
in the range of 16 - 18 $\mu$s for up to 5 qubits. In other cases,
with $\theta_{-}$ in the range of $97-118^{\circ}$, we obtain fidelities
in the range of 0.930 - 0.995, with gate durations of 11 - 23 $\mu$s.
The fidelity can be improved further by increasing the number of MW
pulses.

Since the CNOT gate can be combined with single-qubit gates to yield
a universal set of gates\cite{nielsen,PhysRevLett.102.210502}, the
method presented in this paper represents a universal solution for
implementing quantum computing.

\textit{Conclusion.-} We have experimentally implemented Grover's
quantum search algorithm in a hybrid quantum register in a single
NV center in diamond by indirect control: control pulses were applied
only to the electron spin, which has a much fast response time than
the nuclear spins. In a 2 qubit system, we implemented 4 cases of
the quantum search, in each of which one target state was searched.
The whole procedure for demonstrating the quantum algorithm was implemented,
including the preparation of the pure state, implementation of the
quantum search and reconstruction of the output state. For each target
state, the complete search algorithm was implemented with only 4 MW
pulses. This corresponds to a significant reduction of the control
cost compared with previous works. Further improvements should be
possible by designing the pulse sequence robust against dephasing
effects, or by combining the operations with dynamical decoupling
techniques \cite{PhysRevLett.112.050502,PhysRevLett.115.110502,prar2012,RevModPhys.88.041001}.

\textit{Acknowledgments.-} This project has received funding from
the European Union's Horizon 2020 research and innovation programme
under grant agreement No 828946. The publication reflects the opinion
of the authors; the agency and the commission may not be held responsible
for the information contained in it. We thank Daniel Burgarth for
helpful discussions. 

\bibliographystyle{apsrev}
\bibliography{Grover_resub_arX_Jun26_20}

\section*{Supplemental Material}
\subsection{Setup for optical initialization and detection}

\begin{figure}[h]
\centering 
\includegraphics[width=0.9\columnwidth]{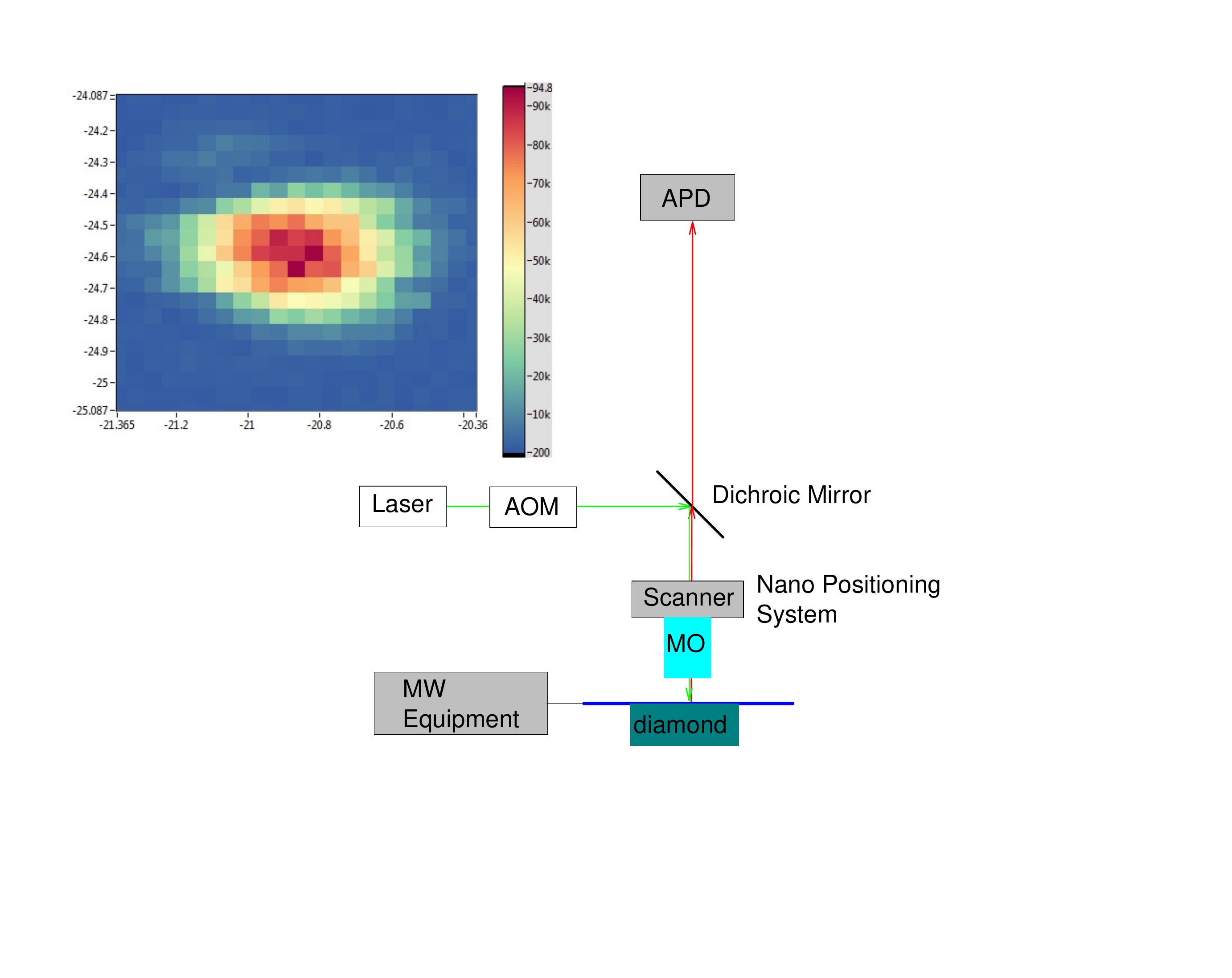} \caption{Schematic of the confocal microscope for initializing and detecting
single NV centers. The inset shows an image of the single NV center
we used in the experiment.}
\label{figresetup} 
\end{figure}

Single NV centers in diamond can be optically addressed, initialized
and detected with a confocal microscope \citep{Suter201750}. In Fig.
\ref{figresetup} we show a schematic of our home-built setup. Here
we used a diode-pumped solid state continuous wave laser with a wavelength
of 532 nm (marked in green in the schematic) for the optical excitation.
For pulsed experiments, an acousto-optical modulator (AOM) with 58
dB extinction ratio and 50 ns rise-time generated the pulses from
the continuous wave laser beam. The microscope objective (MO) lens
was fixed to the nano positioning system that scans the sample in
three dimensions. The fluorescence light with around 637 nm wavelength
(marked in red in the schematic) is also collected by this MO lens,
and passes the dichroic mirror to the avalanche photodiode (APD) detector.
The excitation light is filtered out by the dichroic mirror.

\subsection{System and Hamiltonian}

\begin{figure}[h]
\centering 
\includegraphics[width=0.9\columnwidth]{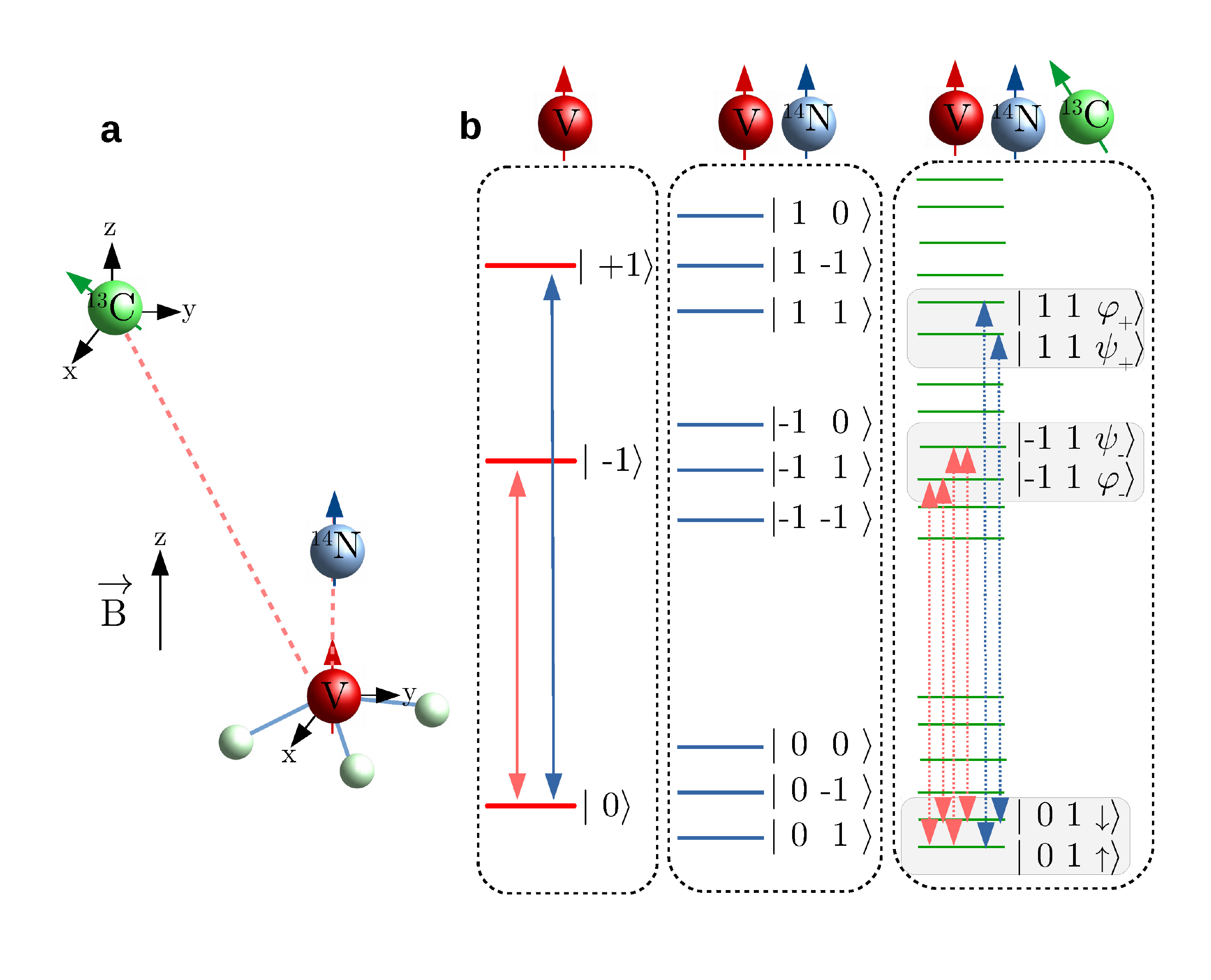} 
 \caption{Characteristics of the system. (a) Structure of the NV system with
the electron spin coupled to one $^{14}$N and one $^{13}$C nuclear
spin. (b) Energy levels of the system consisting of the electron,
$^{14}$N and $^{13}$C nuclear spins. The vertical double arrows
indicate the ESR transitions that we use in this work.}
\label{figreswhole} 
\end{figure}

As illustrated in Figure \ref{figreswhole}, the spin system used
in the present work consists of the electron, $^{14}$N and $^{13}$C
nuclear spins. The static magnetic field $\vec{B}$ is aligned along
the N-V symmetry axis $z$. The relevant Hamiltonian can be written
as 
\begin{eqnarray}
\mathcal{H}/(2\pi) & = & DS_{z}^{2}-\gamma_{e}BS_{z}+P(I_{z}^{N})^{2}-\gamma_{N}BI_{z}^{N}-\gamma_{C}BI_{z}\nonumber \\
 & + & A_{N}S_{z}I_{z}^{N}+A_{zz}S_{z}I_{z}+A_{zx}S_{z}I_{x}.\label{eq:Ham3Spins}
\end{eqnarray}
Here $S_{z}$ and $I_{z}^{N}$ denote the spin-1 operators for the
electron and $^{14}$N spins and $I_{x/z}$ the $^{13}$C spin-1/2
operators. In frequency units, the zero-field splitting is $D=2.87$
GHz, and the $^{14}$N nuclear quadrupole coupling is $P=-4.95$ MHz.
$\gamma_{e/N/C}$ denote the gyromagnetic ratios for the electron,
$^{14}$N and $^{13}$C spins, respectively. $A_{N}=-2.16$ MHz is
the secular part of the hyperfine coupling with the $^{14}$N nuclear
spin~\citep{PhysRevB.89.205202,PhysRevB.47.8816,Yavkin16}, while
$A_{zz}$ and $A_{zx}$ are the relevant components of the $^{13}$C
hyperfine tensor, which are $A_{zz}=-0.152$ MHz and $A_{zx}=0.110$
MHz in the present system.

\begin{figure}[h]
\centering 
\includegraphics[width=1\columnwidth]{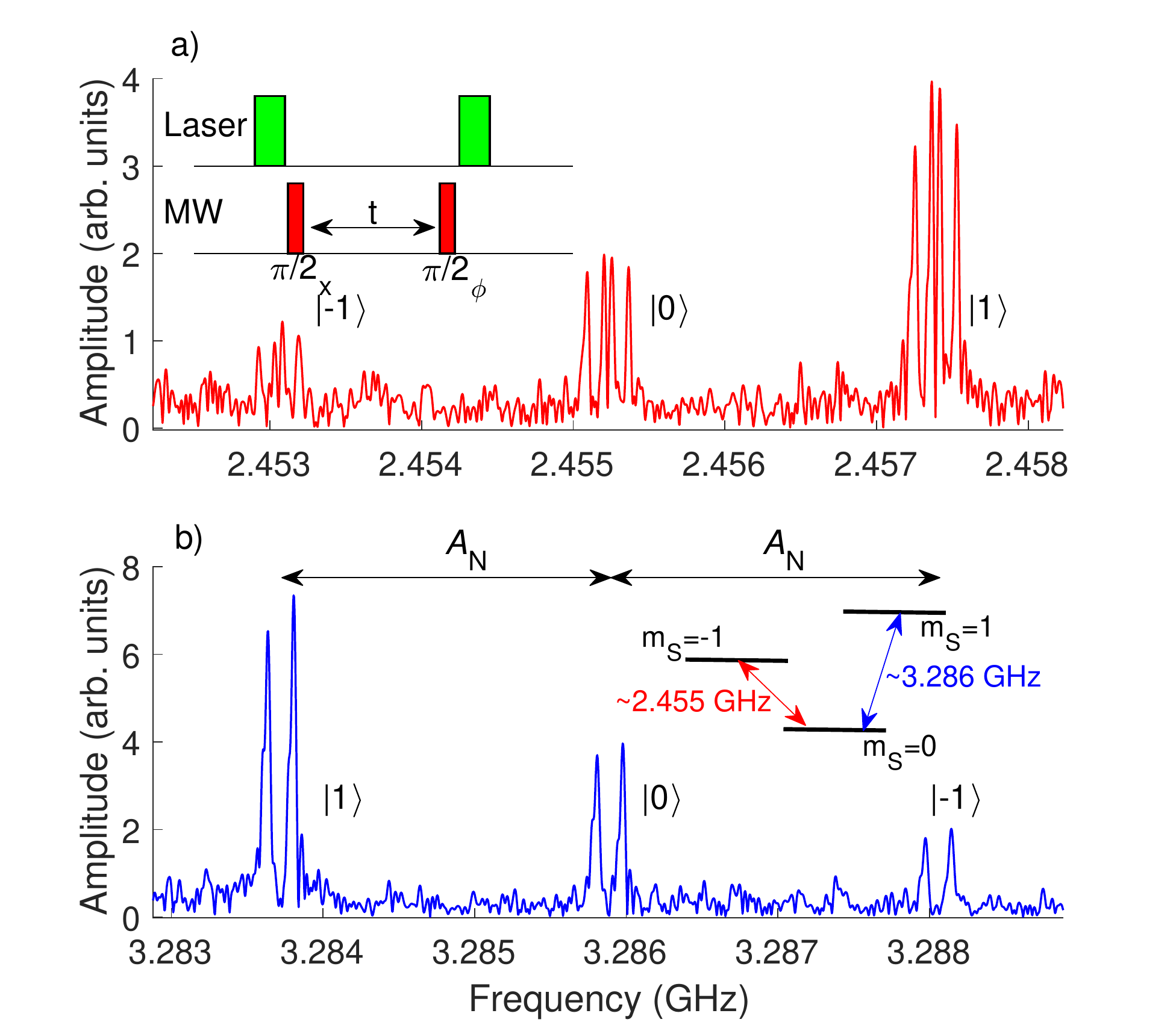} \caption{ESR spectra obtained by ODMR with the pulse sequence {[}inset in (a){]}.
The spectra are obtained in a field of $B=14.8$ mT for the transitions
between (a) $m_{S}=0\leftrightarrow-1$ and (b) $m_{S}=0\leftrightarrow+1$
, as illustrated in the inset in (b){]}. The labels $|0\rangle$,
$|1\rangle$, and $|-1\rangle$ mark the state of the $^{14}$N spin.
The horizontal double arrows indicate the hyperfine coupling with
$^{14}$N.}
\label{figspec} 
\end{figure}

In Figure \ref{figspec} we show the ESR spectra, obtained from a
Free Induction Decay (FID) experiment \citep{Childress281,Ournewpaper}.
The pulse sequence is shown as the inset in Figure \ref{figspec}
(a). The wavelength of the laser pulses is 532 nm, the laser power
$\approx$0.5 mW. The first laser pulse (duration $4$ $\mu$s) initializes
the electron spin into state $m_{S}=0$. The second laser pulse (duration
$0.4$ $\mu$s) is used to detect the population of the state $m_{S}=0$
\citep{Suter201750}.

The MW pulses are on resonance with the transition $m_{S}=0$ $\leftrightarrow$
$m_{S}=-1$ or $m_{S}=0$ $\leftrightarrow$ $m_{S}=1$, for the spectrum
in Figure \ref{figspec} (a) or (b), respectively. In each case, the
pulses have high enough Rabi frequency to cover all the transitions:
in (a), the Rabi frequency is 8.5 MHz, and in (b) 3.7 MHz. The two
MW pulses are both rectangular pulses with flip angle $\pi/2$. The
first pulse generates coherence between states $m_{S}=0$ and $m_{S}=-1$
(a) or $m_{S}=0$ and $m_{S}=1$ (b). After the pulse, the system
evolves under the Hamiltonian (\ref{eq:Ham3Spins}). The second MW
pulse converts the evolved coherence to population, so that it can
be detected by the detection laser pulse. The phase of the second
MW pulse is set as $\phi=2\pi\nu_{d}t$ to generate an effective offset
$\nu_{d}$ in the spectrum, such that all resonance lines appear at
positive frequencies.

If we focus on a subspace where the state of the $^{14}$N is fixed
($m_{N}=1$ in the main text), $\mathcal{H}_{e,C}$ shown in Eq. (1)
in the main text can be diagonalized by the unitary transformation
\begin{equation}
U_{tr}=|1\rangle\langle1|\otimes R_{y}(\theta_{+})+|0\rangle\langle0|\otimes E+|-1\rangle\langle-1|\otimes R_{y}(\theta_{-}),\label{Ut3}
\end{equation}
where $E$ denotes the $2\times2$ identity operator and $R_{y}(\theta_{\pm})=e^{-i\theta_{\pm}I_{y}}$.
The nuclear-spin eigenstates are 
\begin{eqnarray}
|\varphi_{\pm}\rangle & = & |\uparrow\rangle\cos(\theta_{\pm}/2)+|\downarrow\rangle\sin(\theta_{\pm}/2)\nonumber \\
|\psi_{\pm}\rangle & = & -|\uparrow\rangle\sin(\theta_{\pm}/2)+|\downarrow\rangle\cos(\theta_{\pm}/2)\label{eigenC}
\end{eqnarray}
where $\theta_{\pm}=\arctan[A_{zx}/(A_{zz}\mp\nu_{C})]$ denotes the
angle between the nuclear spin quantization axis and the $z$-axis
of our coordinate system for the subsystems where the electron spin
is in $m_{S}=\pm1$. From the experimental spectra, we found $\theta_{-}=87^{\circ}$
and $\theta_{+}=-10^{\circ}$. These results show that the quantization
axis of $^{13}$C for $m_{S}=-1$ ($m_{S}=1$) is approximately perpendicular
to (parallel) to the axis for $m_{S}=0$, respectively, and explain
why the hyperfine splitting from $^{13}$C results in four (two) satellites,
as shown in Figure \ref{figspec} (a-b) \citep{zhang18}.

\subsection{Pulse Sequences\label{subsec:Pulse-Sequences}}

\begin{figure}[b]
\includegraphics[width=1\columnwidth]{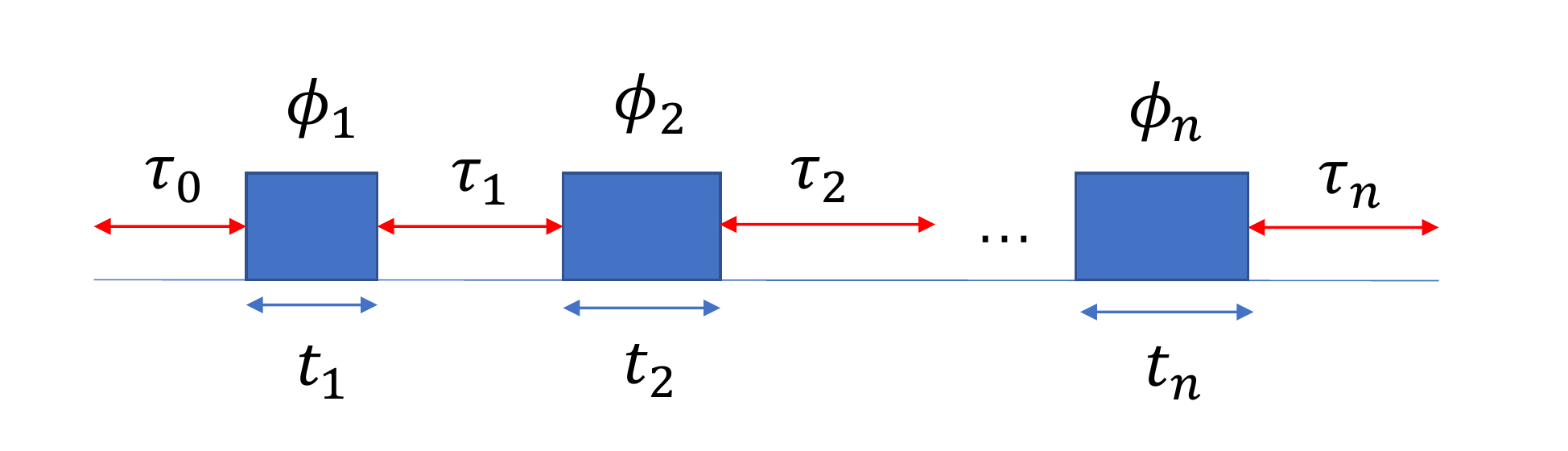} \caption{ Pulse sequence with $n$ pulses and $n+1$ delays for implementing
arbitrary unitary ${\cal U}$. The parameters $t_{i}$ are MW pulse
durations, $\phi_{i}$ are the phases and $\tau_{j}$ the delays.
The amplitude of the pulses is fixed to a Rabi frequency $\omega_{1}/(2\pi)$
= 0.5 MHz, or a value in $[0.48,0.52]$ for the sequence that is robust
against variations in $\omega_{1}$.}
\label{opt_pulse} 
\end{figure}

In a subspace spanned by the states 
\begin{equation}
\{|0\rangle,|-1\rangle\}_{e}\otimes\{|1\rangle\}_{N}\otimes\{|\uparrow\rangle,|\downarrow\rangle\}_{C},
\end{equation}
\label{basissub}the Hamiltonian of the electron-$^{13}$C system
can be represented as 
\begin{equation}
\frac{1}{2\pi}\mathcal{H}_{s}=(-\nu_{C}-\frac{A_{zz}}{2})I_{z}+A_{zz}s_{z}I_{z}+A_{zx}s_{z}I_{x}-\frac{A_{zx}}{2}I_{x}\label{Hsub}
\end{equation}
in the rotating frame \citep{RevModPhys.76.1037} with transform $U_{r}=e^{-i2\pi\nu_{r}ts_{z}}$
where $\nu_{r}=D+\nu_{e}-A_{N}$. Here $s_{z}$ denotes the pseudo-spin
1/2 operator for the electron spin in the subspace and $\nu_{C}=\gamma_{C}B$.

The pulse sequence to implement an arbitrary target unitary ${\cal {U}}$
is shown as Figure \ref{opt_pulse}, where $n$ MW pulses with fixed
Rabi frequency $\omega_{1}$ and $n+1$ delays are used. The propagators
for the individual MW pulses can be written as $U_{k}^{MW}=e^{-i\mathcal{H}_{k}^{MW}t_{k}}$
where $\mathcal{H}_{k}^{MW}=\mathcal{H}_{s}+\omega_{1}[s_{x}\cos(\phi_{k})+s_{y}\sin(\phi_{k})]$,
with $k=1,\cdots,n$, and for the free evolutions as $U_{k}^{d}=e^{-i\mathcal{H}_{s}\tau_{k}}$,
with $k=0,\cdots,n$. The total unitary ${\cal U}$ is a time ordered
product of the $U_{k}^{MW}$ and $U_{k}^{d}$, and is a function of
the pulse parameters $(t_{1},\cdots,t_{n},\phi_{1},\cdots,\phi_{n},\tau_{0},\cdots,\tau_{n})$.
The goal is to design a unitary ${\cal U}$ that maximizes the fidelity
\begin{equation}
F_{g}=|\mathrm{Tr(U_{T}^{\dagger}{\cal U}})|/4\label{Fgate}
\end{equation}
where $U_{T}$ denotes the target unitary operation. We used a genetic
algorithm for the numerical search to obtain an optimal set of parameters.

\begin{figure}[b]
\includegraphics[width=1\columnwidth]{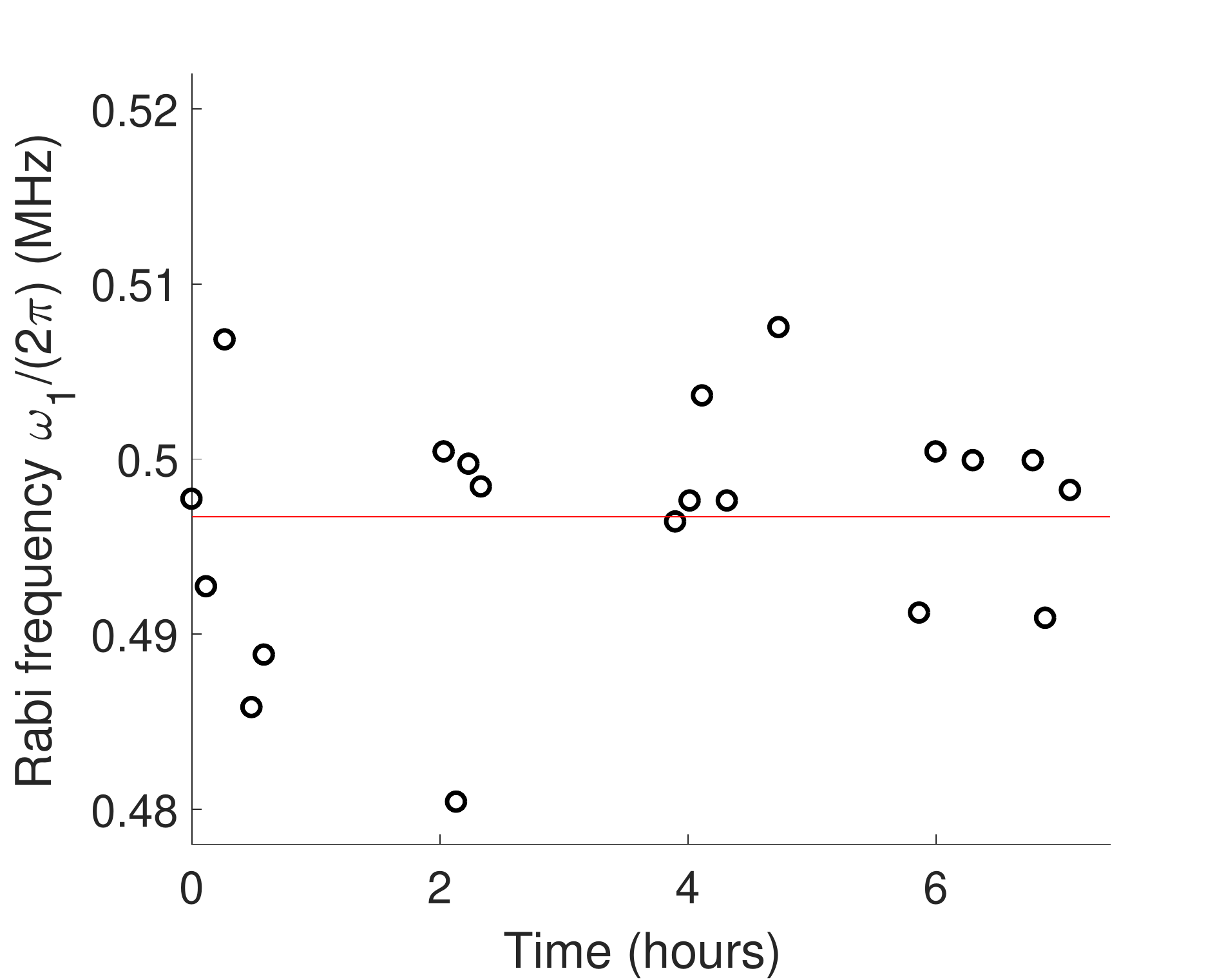}\caption{Variation of the Rabi frequency $\omega_{1}/(2\pi)$ with time. The
red line indicates the average. }
\label{Rabifluc}
\end{figure}

In the experiment of the Grover's quantum search, we use one pulse
sequence with $4$ MW pulses and 5 delays to implement the circuit
shown in Figure 1 (c) in the main text for each target state. The
effect of the individual MW pulses is sensitive to variations in $\omega_{1}$.
In Figure \ref{Rabifluc}, we illustrate the fluctuation of $\omega_{1}/(2\pi)$
over 7 hours. The fluctuation of MW power could be mainly attributed
to the amplifiers in the circuit, where we used three amplifiers in
series from mini circuits {[}models ZHL-16W-43-S+ (power amplifier),
ZFL-500LN+ (pre-amplifier) and ZX60-4016E-S+ (pre-amplifier){]}.

As shown in Figure \ref{Rabifluc} the MW field strength (Rabi frequency)
varies by about 0.03 MHz over the time scale of the experiment (3
hours). To obtain good fidelity in experiments where the actual MW
amplitude deviates from the ideal one, we optimized the pulse sequences
for a range of amplitudes, taking the average fidelity as the performance
measure, so that the resulting sequences are robust against the amplitude
fluctuation. We used the range $\omega_{1}/2\pi=[0.48,0.52]$ MHz,
which covers the observed range of amplitudes, as shown in Figure
\ref{Rabifluc}. The theoretical fidelities for the targets $|00\rangle$,
$|01\rangle$, $|10\rangle$ and $|11\rangle$ are $98.2\%$, $97.9\%$,
$98.0\%$ and $97.1\%$, respectively. The parameters of the pulse
sequences are listed in Tables \ref{tabparam00} - \ref{tabparam11}.

\begin{table*}
\begin{subtable}{0.45\textwidth} \centering %
\begin{tabular}[t]{|c|c|c|}
\hline 
Delay ($\mu$s)  & Pulse duration ($\mu$s)  & Phase ($^{\circ}$) \tabularnewline
\hline 
$\tau_{0}=0.987$  &  & \tabularnewline
\hline 
$\tau_{1}=1.968$  & $t_{1}=0.976$  & $\phi_{1}=261$ \tabularnewline
\hline 
$\tau_{2}=2.418$  & $t_{2}=0.510$  & $\phi_{2}=213$ \tabularnewline
\hline 
$\tau_{3}=2.465$  & $t_{3}=0.394$  & $\phi_{3}=141$ \tabularnewline
\hline 
$\tau_{4}=1.136$  & $t_{4}=1.104$  & $\phi_{4}=90$ \tabularnewline
\hline 
\end{tabular}\caption{Target $|00\rangle$.}
\label{tabparam00} \end{subtable} 
\begin{subtable}{0.45\textwidth} \centering %
\begin{tabular}{|c|c|c|}
\hline 
Delay ($\mu$s)  & Pulse duration ($\mu$s)  & Phase ($^{\circ}$) \tabularnewline
\hline 
$\tau_{0}=0.559$  &  & \tabularnewline
\hline 
$\tau_{1}=0.399$  & $t_{1}=0.555$  & $\phi_{1}=302$ \tabularnewline
\hline 
$\tau_{2}=2.905$  & $t_{2}=1.290$  & $\phi_{2}=195$ \tabularnewline
\hline 
$\tau_{3}=2.476$  & $t_{3}=0.472$  & $\phi_{3}=196$ \tabularnewline
\hline 
$\tau_{4}=1.139$  & $t_{4}=1.423$  & $\phi_{4}=90$ \tabularnewline
\hline 
\end{tabular}\caption{Target $|01\rangle$.}
\label{tabparam01} \end{subtable} 

\bigskip{}
\begin{subtable}{0.45\textwidth} \centering %
\begin{tabular}{|c|c|c|}
\hline 
Delay ($\mu$s)  & Pulse duration ($\mu$s)  & Phase ($^{\circ}$) \tabularnewline
\hline 
$\tau_{0}=0.995$  &  & \tabularnewline
\hline 
$\tau_{1}=2.518$  & $t_{1}=1.484$  & $\phi_{1}=102$ \tabularnewline
\hline 
$\tau_{2}=2.353$  & $t_{2}=0.542$  & $\phi_{2}=340$ \tabularnewline
\hline 
$\tau_{3}=0.361$  & $t_{3}=0.210$  & $\phi_{3}=47$ \tabularnewline
\hline 
$\tau_{4}=1.191$  & $t_{4}=1.262$  & $\phi_{4}=90$ \tabularnewline
\hline 
\end{tabular}\caption{Target $|10\rangle$.}
\label{tabparam10} \end{subtable}
\begin{subtable}{0.45\textwidth} \centering %
\begin{tabular}{|c|c|c|}
\hline 
Delay ($\mu$s)  & Pulse duration ($\mu$s)  & Phase ($^{\circ}$) \tabularnewline
\hline 
$\tau_{0}=1.892$  &  & \tabularnewline
\hline 
$\tau_{1}=2.345$  & $t_{1}=0.995$  & $\phi_{1}=198$ \tabularnewline
\hline 
$\tau_{2}=2.583$  & $t_{2}=0.541$  & $\phi_{2}=0$ \tabularnewline
\hline 
$\tau_{3}=2.576$  & $t_{3}=0.452$  & $\phi_{3}=90$ \tabularnewline
\hline 
$\tau_{4}=0.665$  & $t_{4}=0.939$  & $\phi_{4}=90$ \tabularnewline
\hline 
\end{tabular}\caption{Target $|11\rangle$.}
\label{tabparam11} \end{subtable}\caption{Parameters for the pulse sequence robust against the fluctuation of
the MW amplitude to implement Grover's search with various target
states. }
\label{tabparamrobust} 
\end{table*}

\subsection{Error analysis}

Based on the measured density matrices shown in Figure 3 in the main
text, we evaluate the performance of the quantum search in the following
way. 
\begin{enumerate}
\item As listed in the main text, the experimental fidelity of the initial
state $\rho_{exp}^{ini}$ is $F_{ini}=0.92$. The loss of fidelity
can be mainly attributed to the imperfection of the tomography. Examples
include the non-negligible off-diagonal elements in the density matrix,
which reflect imperfect tomography. By applying the ideal circuit
of Figure 1(c) in the main text to $\rho_{exp}^{ini}$, we obtain
$\rho_{1}$, and then calculate the state fidelity $F_{1}=Tr\{\rho_{th}\rho_{1}\}=0.925$,
where $\rho_{th}=|11\rangle\langle11|,$ which is close to $F_{ini}$.
Since we used the same tomography procedure for the initial and the
final state, we conclude that the main contribution to the observed
infidelity originates from the tomographic analysis. 
\item By applying the theoretical MW pulse sequence to the ideal initial
state $\rho^{ini}=|00\rangle\langle00|$, we obtain $\rho_{2}=U_{MW}\rho^{ini}U_{MW}^{\dagger}$,
and the fidelity $F_{2}=Tr\{\rho_{th}\rho_{2}\}=0.967.$ 
\item Combining the measured fidelity $F_{|11\rangle}=0.85$ for Grover's
search obtained from the results shown in Figure 3 in the main text,
we estimate the fidelity of the implemented quantum search as $F_{3}=F_{|11\rangle}/(F_{1}F_{2})=0.95$.
In our system, we can treat the decoherence time of the electron spin
between $T_{2}^{*}\approx35\mu$s measured from the ESR FID signal
and $T_{2}\approx1$ ms from the dynamical decoupling sequence \citep{PhysRevA.98.052354}.
The duration of the pulse sequence is 12.989 $\mu$s, which is not
negligible compared to $T_{2}^{*}$. Therefore the decoherence appears
to be the main contibution to the error in the search. Increasing
the coherence time of the electron spin, e.g., by decreasing the concentration
of substitutional nitrogen spins in the diamod sample, should further
improve the experiment.

\subsection{Effects of $^{14}$N in Grover's search}

The MW pulse sequences for implementing Grover's search targeted the
subspace of $^{14}$N state $m_{N}=1$, where $m_{N}$ denotes magnetic
quantum number for $^{14}$N spin. We here simulated the pulse sequences
applied to the whole space of $^{14}$N as $m_{N}=\{0,-1,1\}$ to
investigate the quality of selectivity for the subspace of $m_{N}=1$.
To simplify the procedure, we here only consider the electron and
the $^{14}$N spins \citep{Ournewpaper}. For our system, the initial
state can be represented as \citep{zhang18} 
\begin{equation}
\rho_{0}^{e,N}=|0\rangle\langle0|_{e}\otimes\left(\sum c_{m_{N}}|m_{N}\rangle\langle m_{N}|\right)\label{eq:inien}
\end{equation}
where $c_{m_{N}}\approx4/7$, $2/7$, $1/7$, for $m_{N}=1$, $0$,
$-1$ \citep{zhang18}. In the ideal case, the MW pulses that are
set to select the subspace of $m_{N}=1$ do not change the populations
of the states $|0,0\rangle$ and $|0,-1\rangle$ in the initial state
(\ref{eq:inien}).

After applying the pulse sequence to the initial state $\rho_{0}^{e,N}$,
we measured the populations of states $|0,0\rangle$ and $|0,-1\rangle$,
and found that they decreased from the initial values.

In the procedure of state tomography in experiment, we extracted the
population of the state $|0,1\rangle$ by subtracting initial populations
of states $|0,0\rangle$ and $|0,-1\rangle$ {[}given by $c_{0}$
and $c_{-1}$ in Eq. (\ref{eq:inien}){]} from the total population
of the bright electron state $m_{S}=0$. Therefore the changes of
the populations of states $|0,0\rangle$ and $|0,-1\rangle$ in Grover's
search lead to a loss of measured population of the computational
subspace of $m_{N}=1$. In Table \ref{tablossP}, we list the loss
of population in Grover's search for each target state. 
\begin{table}
\begin{tabular}{|c|c|}
\hline 
Target state in Grover's search  & Loss of the population\tabularnewline
\hline 
$|00\rangle$  & 0.025\tabularnewline
\hline 
$|01\rangle$  & 0.0068\tabularnewline
\hline 
$|10\rangle$  & 0.020\tabularnewline
\hline 
$|11\rangle$  & 0.027\tabularnewline
\hline 
\end{tabular}

\caption{The loss of the population of the computational subspace in Grover's
search due to effects of the $^{14}$N coupled to the electron spin. }
\label{tablossP} 
\end{table}

We use numerical simulation to investigate the dependence of the loss
of the population $L_{p}$ on the polarization of the $^{14}$N spin
$p_{N}$ using the case of the target state $|11\rangle$ as an example.
The input state is 
\begin{eqnarray}
\rho_{0,sim}^{e,N}=|0\rangle\langle0|_{e} & \otimes & (p_{N}|1\rangle\langle1|+\frac{1-p_{N}}{2}|0\rangle\langle0|\nonumber \\
 & + & \frac{1-p_{N}}{2}|-1\rangle\langle-1|)\label{eq:iniensim}
\end{eqnarray}
The pulse sequence in simulation is shown in Figure \ref{opt_pulse},
with the parameters listed in Table \ref{tabparam11}. We calculated
the sum of the populations in the states $|m_{S}=-1,m_{N}=0\rangle$
and $|m_{S}=-1,m_{N}=-1\rangle$ as the loss of the population $L_{p}$
from the computational space. Figure \ref{LossP} shows the result.

\begin{figure}[b]
\includegraphics[width=1\columnwidth]{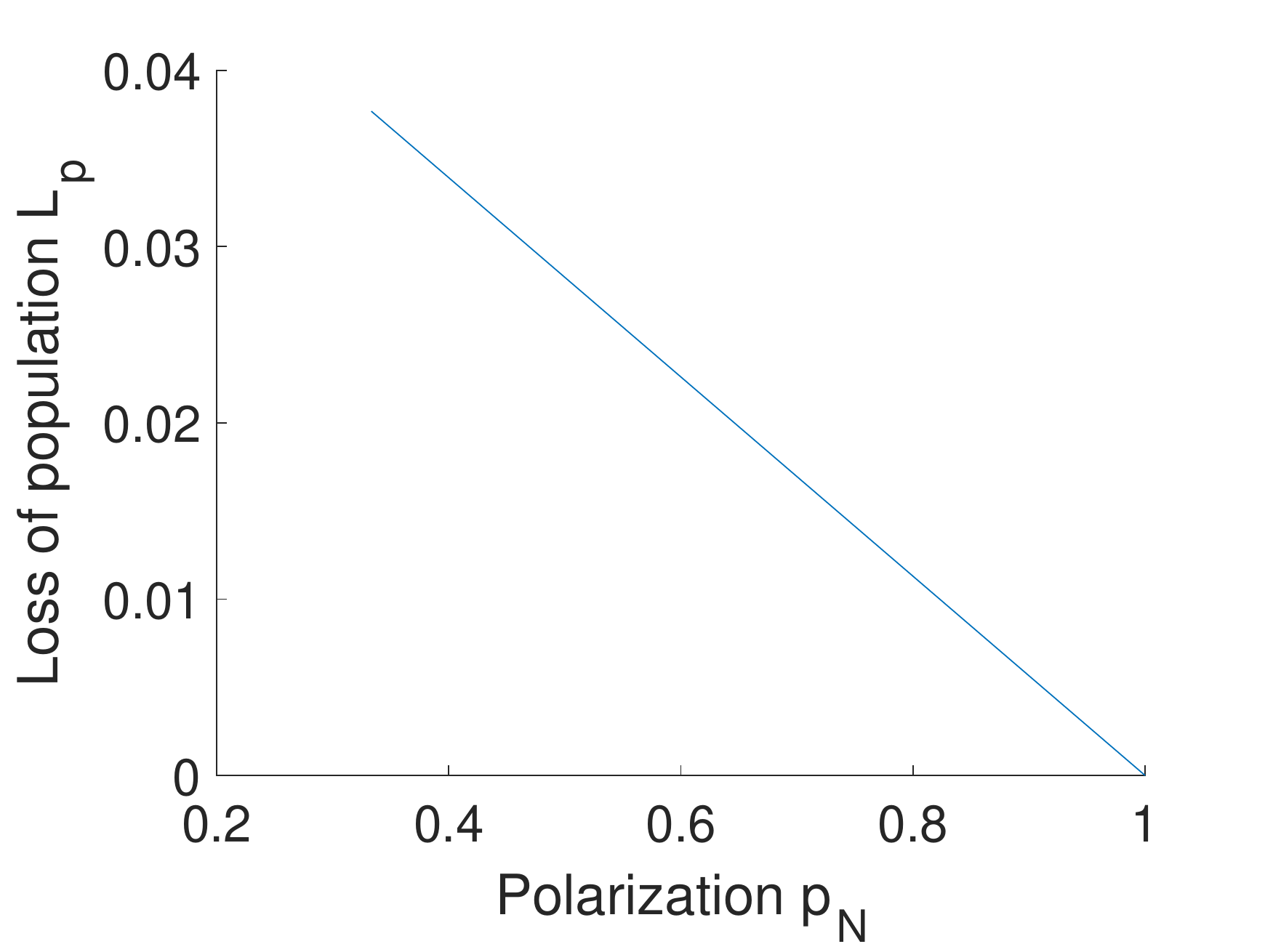} \caption{Dependence of the loss of population $L_{p}$ on the initial polarization
of the $^{14}$N spin $p_{N}$. The range of $p_{N}$ was chosen from
$1/3$ to $1$, corresponding to the $^{14}$N in the maximal mixed
( or identity) state and pure state with $m_{N}=1$.}
\label{LossP} 
\end{figure}

\subsection{Pulse sequence for pure state preparation}

\begin{figure}[b]
\includegraphics[width=1\columnwidth]{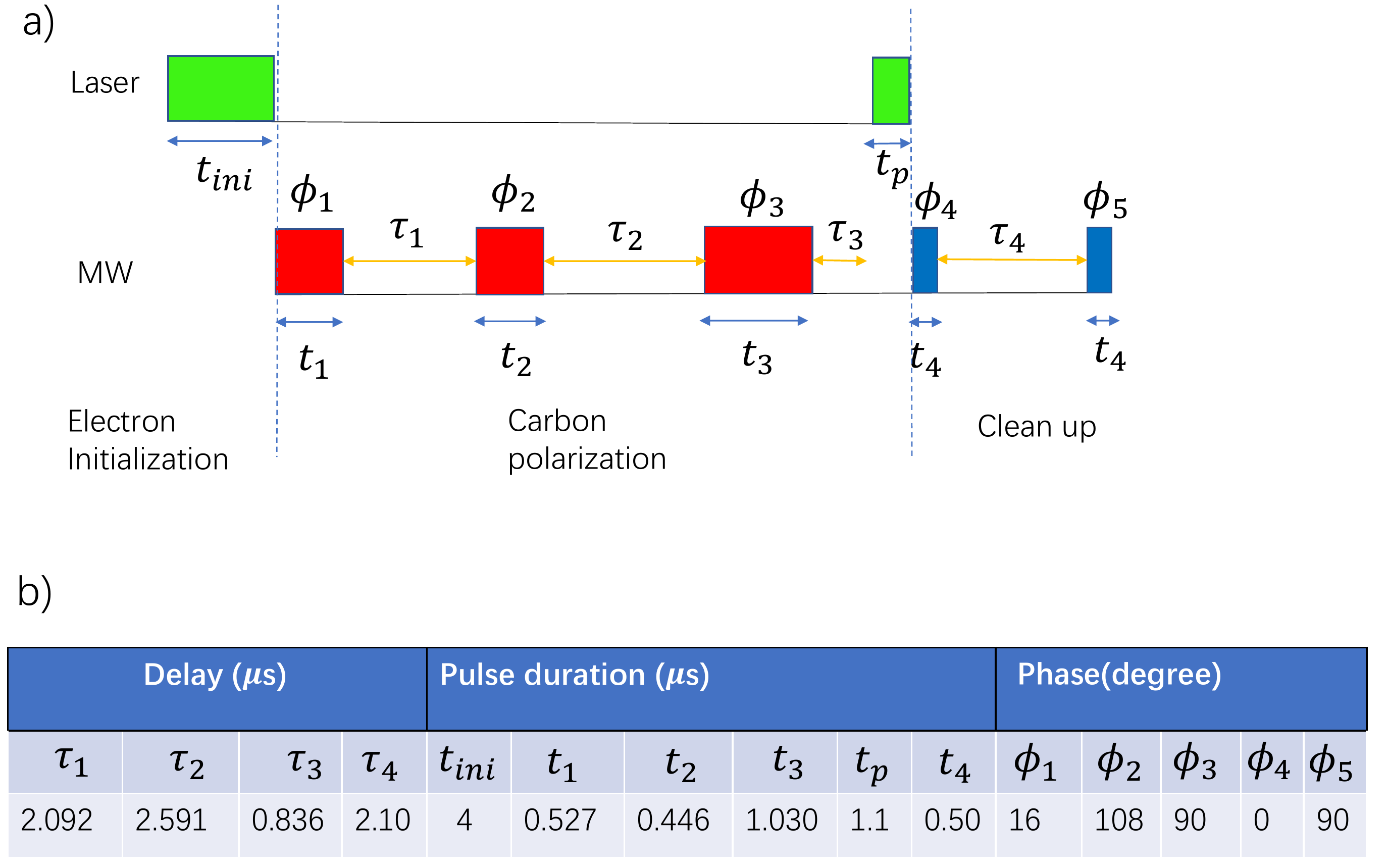} \caption{Pulse sequence (a) with the parameters listed in table (b) for preparing
the pure state $|00\rangle$. The amplitude of the pulses is fixed
to a Rabi frequency $\omega_{1}/(2\pi)$ = 0.5 MHz. The MW pulses
indicated in red rectangles are on resonance with the transition frequency
between $m_{S}=0$ and $-1$, those in blue with $m_{S}=0$ $\leftrightarrow$
$1$. The clean up step is a CNOT-like gate \citep{Ournewpaper} that
moves the leftover (undesired) population of state $|01\rangle$ out
of the computational space.}
\label{pure_pulse} 
\end{figure}

Figure \ref{pure_pulse} (a) shows the pulse sequence to prepare the
pure state $|00\rangle$, and (b) the parameters \citep{zhang18}.
In the initialization step, the electron is set to $m_{S}=0$, while
the $^{13}$C spin is in the maximally mixed state. The $^{13}$C
spin can be polarized by the MW and laser pulses indicated in the
step of carbon polarization. The state of the two spins after the
second laser pulse can be represented as 
\begin{equation}
\rho_{p}=|0\rangle\langle0|\otimes[p|0\rangle\langle0|+(1-p)|1\rangle\langle1|+c(|0\rangle\langle1|+|1\rangle\langle0|)]\label{rhopol}
\end{equation}
with $p$ measured as 0.83, and c as 0.08. The state $\rho_{p}$ can
be further purified by moving the leftover population of state $|01\rangle$
and the coherence out of the space for quantum computing \citep{swathi19}.
As a result, we obtain a pure state $\rho^{ini}=|00\rangle\langle00|$
as the initial state for the quantum search, after re-normalizing
the total population in the two qubit computational space to unity.

\subsection{Additional data for optimal control}

\subsubsection{Parameters of the pulse sequence with fixed Rabi of $0.5$ MHz}

If we assume that the amplitude of the MW pulses can be exactly controlled,
e.g., the Rabi frequency $\omega_{1}$ is fixed to $0.5$ MHz, we
can remove the condition of robustness against fluctuations of $\omega_{1}$
from the optimization of the parameters, and therefore we can increase
the theoretical fidelity of the pulse sequence. We list the resulting
parameters in Tables \ref{tabparam00fix} -\ref{tabparam11fix}. The
average fidelity for the four target states is $0.988$, slightly
higher than in the robust case, which is $0.980$, obtained from the
parameters listed in Tables \ref{tabparam00} -\ref{tabparam11}.
The average duration of the pulse sequences for the four target states
is $12.2$ $\mu$s, similar to the robust case where it was $11.8$
$\mu$s.

If we use 6 MW pulses instead of 4, we can improve the fidelity in
the case of target state $|01\rangle$ to $0.995$, with a sequence
duration of $19.23$ $\mu$s. The parameters are llisted in Table
\ref{tabparam01_6}.

\begin{table*}
\begin{subtable}{0.45\textwidth} \centering %
\begin{tabular}{|c|c|c|}
\hline 
Delay ($\mu$s)  & Pulse duration ($\mu$s)  & Phase ($^{\circ}$) \tabularnewline
\hline 
$\tau_{0}=0.944$  &  & \tabularnewline
\hline 
$\tau_{1}=1.922$  & $t_{1}=1.139$  & $\phi_{1}=266$ \tabularnewline
\hline 
$\tau_{2}=2.554$  & $t_{2}=0.481$  & $\phi_{2}=201$ \tabularnewline
\hline 
$\tau_{3}=1.802$  & $t_{3}=0.402$  & $\phi_{3}=142$ \tabularnewline
\hline 
$\tau_{4}=0.777$  & $t_{4}=1.126$  & $\phi_{4}=90$ \tabularnewline
\hline 
\end{tabular}\caption{Target state $|00\rangle$, with theoretical fidelity 0.991.}
\label{tabparam00fix} \end{subtable} 
\begin{subtable}{0.45\textwidth} \centering %
\begin{tabular}{|c|c|c|}
\hline 
Delay ($\mu$s)  & Pulse duration ($\mu$s)  & Phase ($^{\circ}$) \tabularnewline
\hline 
$\tau_{0}=1.099$  &  & \tabularnewline
\hline 
$\tau_{1}=0.608$  & $t_{1}=0.479$  & $\phi_{1}=285$ \tabularnewline
\hline 
$\tau_{2}=2.442$  & $t_{2}=0.881$  & $\phi_{2}=2$ \tabularnewline
\hline 
$\tau_{3}=2.993$  & $t_{3}=0.608$  & $\phi_{3}=197$ \tabularnewline
\hline 
$\tau_{4}=1.768$  & $t_{4}=1.323$  & $\phi_{4}=90$ \tabularnewline
\hline 
\end{tabular}\caption{Target state $|01\rangle$, with theoretical fidelity 0.984. }
\label{tabparam01fix} \end{subtable} 

\bigskip{}
\begin{subtable}{0.45\textwidth} \centering %
\begin{tabular}{|c|c|c|}
\hline 
Delay ($\mu$s)  & Pulse duration ($\mu$s)  & Phase ($^{\circ}$) \tabularnewline
\hline 
$\tau_{0}=0.634$  &  & \tabularnewline
\hline 
$\tau_{1}=1.763$  & $t_{1}=1.698$  & $\phi_{1}=112$ \tabularnewline
\hline 
$\tau_{2}=1.603$  & $t_{2}=0.448$  & $\phi_{2}=313$ \tabularnewline
\hline 
$\tau_{3}=1.945$  & $t_{3}=0.426$  & $\phi_{3}=23$ \tabularnewline
\hline 
$\tau_{4}=1.261$  & $t_{4}=1.224$  & $\phi_{4}=90$ \tabularnewline
\hline 
\end{tabular}\caption{Target state $|10\rangle$, with theoretical fidelity 0.990. }
\label{tabparam10fix} \end{subtable} 
\begin{subtable}{0.45\textwidth} \centering %
\begin{tabular}{|c|c|c|}
\hline 
Delay ($\mu$s)  & Pulse duration ($\mu$s)  & Phase ($^{\circ}$) \tabularnewline
\hline 
$\tau_{0}=1.751$  &  & \tabularnewline
\hline 
$\tau_{1}=2.439$  & $t_{1}=1.069$  & $\phi_{1}=10$ \tabularnewline
\hline 
$\tau_{2}=1.661$  & $t_{2}=1.584$  & $\phi_{2}=125$ \tabularnewline
\hline 
$\tau_{3}=3.255$  & $t_{3}=0.514$  & $\phi_{3}=51$ \tabularnewline
\hline 
$\tau_{4}=1.183$  & $t_{4}=0.858$  & $\phi_{4}=90$ \tabularnewline
\hline 
\end{tabular}\caption{Target state $|11\rangle$, with theoretical fidelity 0.990. }
\label{tabparam11fix} \end{subtable} \caption{Parameters of the non-robust pulse sequences to implement Grover's
search with various target states. }
\label{tabparamfix} 
\end{table*}

\begin{table}
\begin{tabular}{|c|c|c|}
\hline 
Delay ($\mu$s)  & Pulse duration ($\mu$s)  & Phase ($^{\circ}$) \tabularnewline
\hline 
$\tau_{0}=0.887$  &  & \tabularnewline
\hline 
$\tau_{1}=0.834$  & $t_{1}=1.305$  & $\phi_{1}=192$ \tabularnewline
\hline 
$\tau_{2}=2.994$  & $t_{2}=1.570$  & $\phi_{2}=46$ \tabularnewline
\hline 
$\tau_{3}=1.994$  & $t_{3}=1.528$  & $\phi_{3}=326$ \tabularnewline
\hline 
$\tau_{4}=1.734$  & $t_{4}=0.770$  & $\phi_{4}=54$ \tabularnewline
\hline 
$\tau_{5}=1.204$  & $t_{5}=0.709$  & $\phi_{5}=238$ \tabularnewline
\hline 
$\tau_{6}=2.598$  & $t_{6}=1.103$  & $\phi_{6}=90$ \tabularnewline
\hline 
\end{tabular}\caption{Parameters for the pulse sequence to implement Grover's search with
target state $|01\rangle$ with 6 pulses. The theoretical fidelity
is 0.995. }
\label{tabparam01_6} 
\end{table}

\subsubsection{Number of MW pulses}

We use the optimization of the pulse sequence for the Grover's search
with target state as $|01\rangle$ as an example to illustrate how
to choose the number of the MW pulses in the optimal control. Figure
\ref{pure_pulse-1} shows the obtained fidelities and gate durations
when we used 3, 4 and 6 pulses, respectively. As expected, the fidelity
increases with the number of pulses. However, it also leads to longer
gate duration (except the case from 3-4 pulses, but the fidelity of
3 pulses is not high enough), and therefore enhance the effects of
dephasing. Moreover, more pulses also result in more operation errors
in experimental implementation. We chose 4 puslse as a compromise
for the implementation of the Grover's search.

\begin{figure}[b]
\includegraphics[width=1\columnwidth]{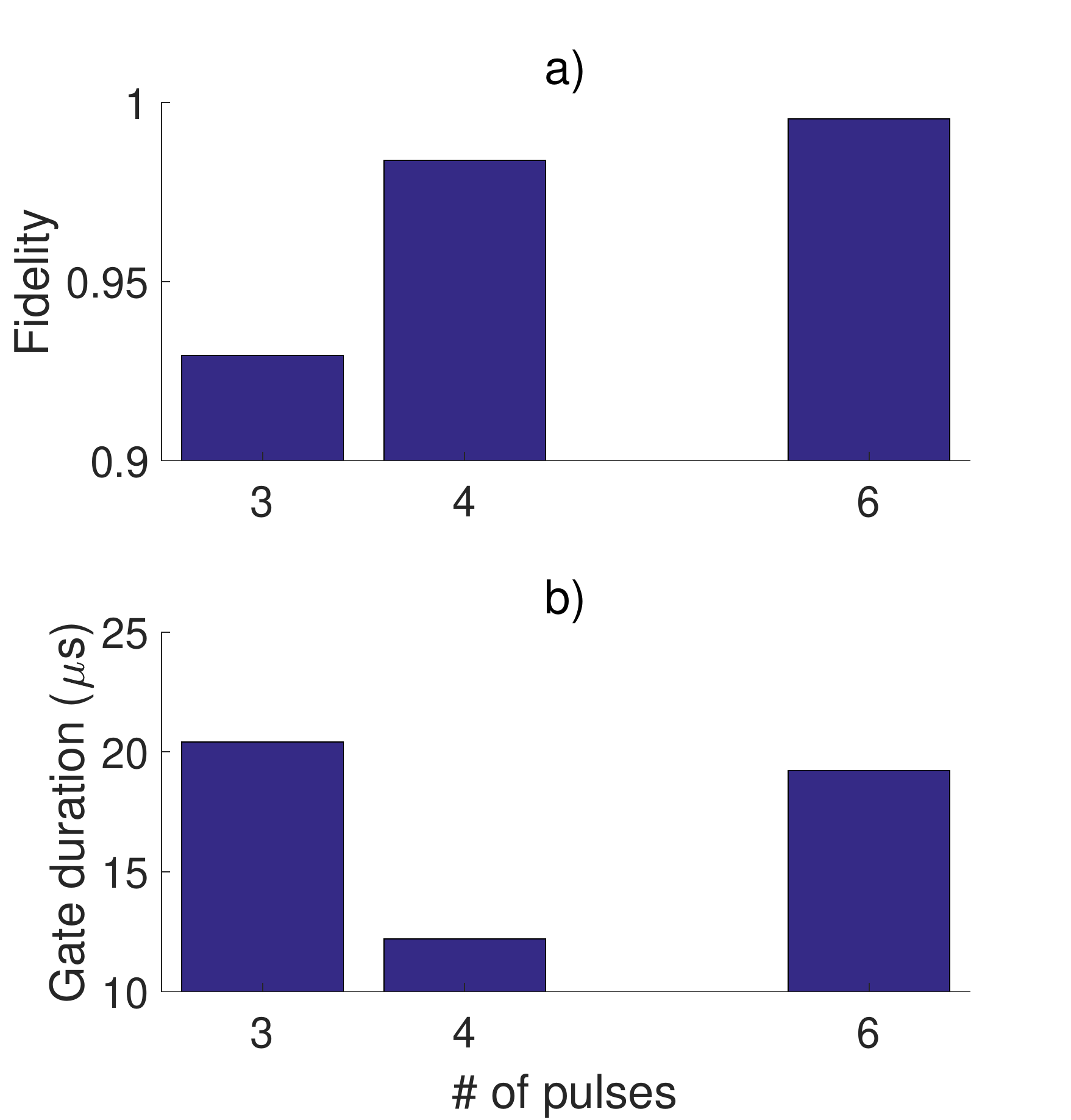} \caption{Dependence of the fidelity (a) and gate duration (b) on the number
of the MW pulses in the optimal control for the Grover's search for
the target state $|01\rangle$.}
\label{pure_pulse-1}
\end{figure}

\subsubsection{Comparison of optimization for CNOT-like gate and quantum search}

We compare the optimization of the pulse sequences for two unitaries.
One unitary is chosen as a CNOT-like gate, where the electron spin
acts as the control qubit, and $^{13}$C spin as the target. The operation
for the target qubit is a rotation $R_{x}(\pi)=e^{-i\pi I_{x}}$ .
The other unitary is Grover's search for the target state $|11\rangle$
shown in Figure 1 (c). To clarify this circuit consisting of two CNOT-like
gates (instead of CNOT gates) and 5 more single qubit gates, we represented
this circuit as Figure \ref{timesc} (a), where the dash-dotted rectangles
indicate CNOT-like gates.

In Figures \ref{timesc} (b-c), we illustrate the decrease of the
infidelity $1-F_{g}$ (or the increase of the gate fidelity $F_{g}$),
with the generations (or iterations) in the optimization. It shows
that the optimization for the CNOT-like gate is much faster than for
Grover's search which is more compex than the CNOT-like gate. The
parameters are listed in Tables \ref{tabparamcont} and \ref{tabparam11fix},
respectively.

\begin{figure}[b]
\includegraphics[width=1\columnwidth]{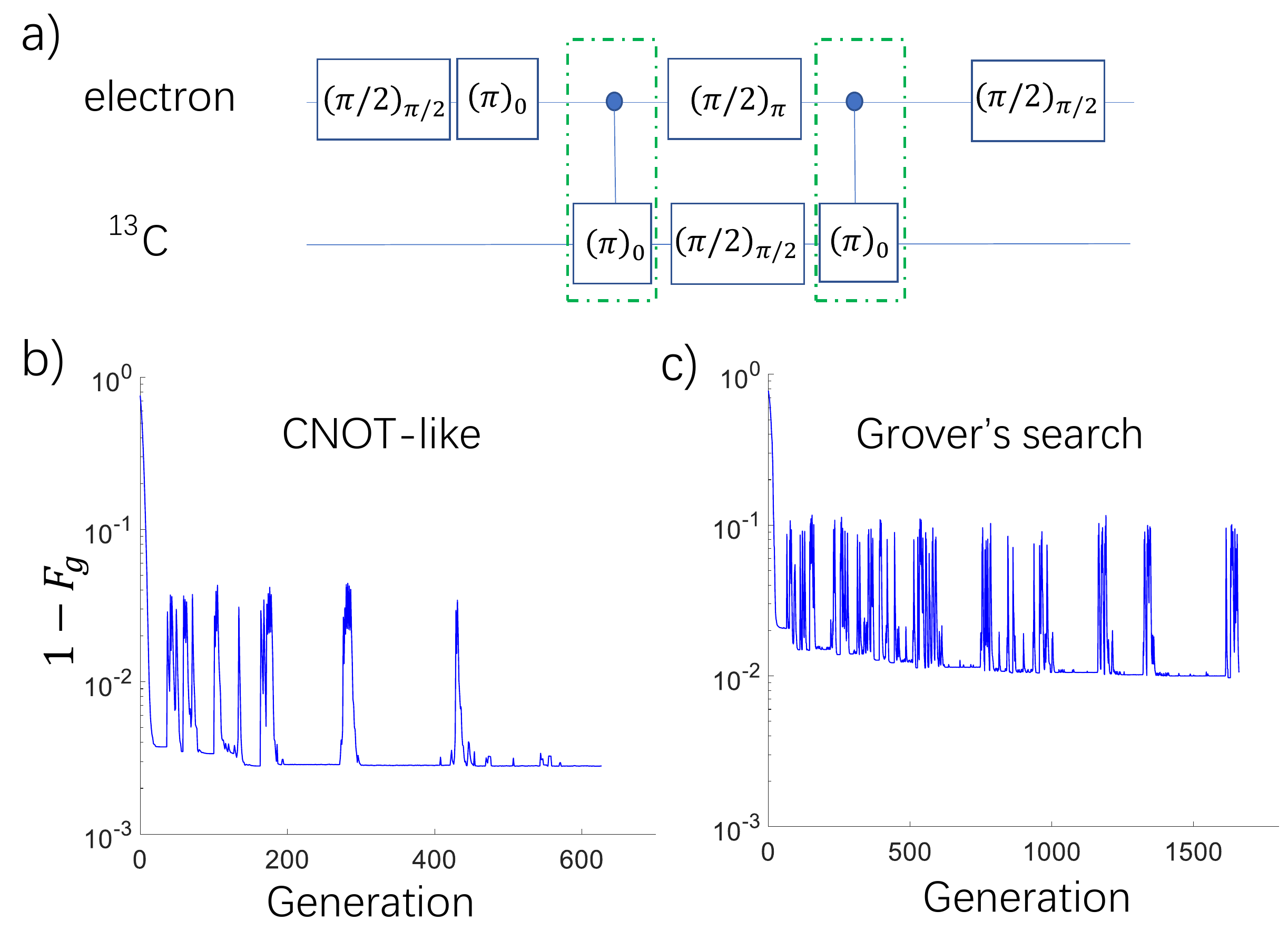} \caption{(a) The simplified circuit for implementing Grover's search with target
state $|11\rangle$, identical to the circuit in Figure 1 (c) in the
main text. In figure (a), the operations indicated in the solid rectangles
denote single-qubit rotations $(\theta)_{\phi}=e^{-i\theta[I_{x}\cos(\phi)+I_{y}\sin(\phi)]}$
or $e^{-i\theta[s_{x}\cos(\phi)+s_{y}\sin(\phi)]}$, applied to $^{13}$C
(bottom line) or electron (top line) spin, respectively. The dash-dotted
rectangles indicate the CNOT-like gates. (b-c) The minimization of
$1-F_{g}$ by the genetic algorithm with the generations during the
optimization process for the CNOT-like gate and Grover's quantum search
with target state $|11\rangle$, as indicated in the panels, respectively.
The gate fidelity $F_{g}$ is $>0.997$ for the CNOT-like gate, and
$>0.990$ for the quantum search.}
\label{timesc} 
\end{figure}

\subsection{Simulation of multiple qubits system}

\begin{figure}[b]
\includegraphics[width=1\columnwidth]{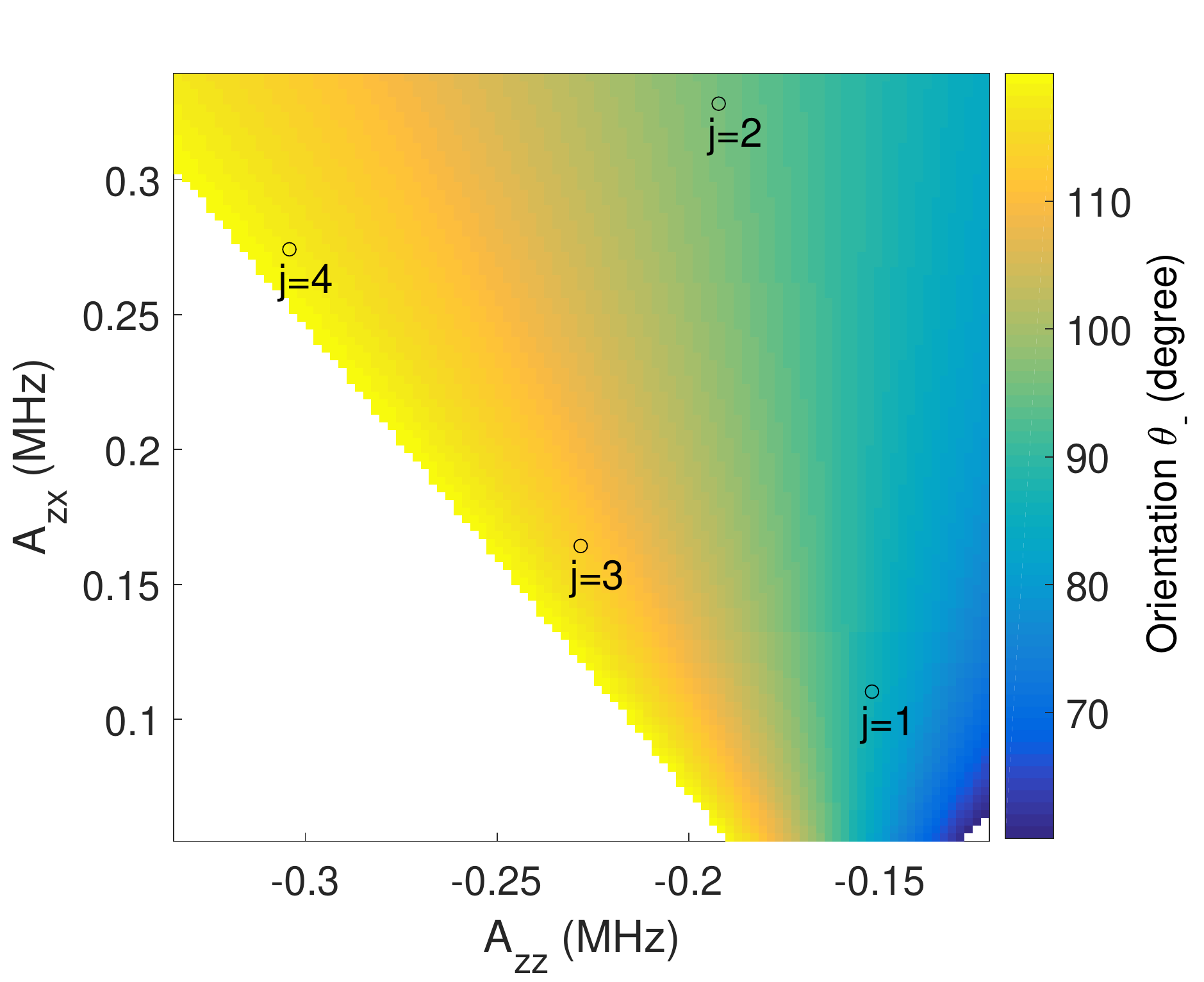} \caption{The orientation of the quantization axis in the subspace of electron
state $m_{S}=-1$ as a function of the coupling constants $A_{zz}$and
$A_{zx}$.}
\label{quancoup} 
\end{figure}

By generalizing the Hamiltonian (\ref{Hsub}) for the two spin system,
we can write the Hamiltonian of one electron and $n$- $^{13}$C spins
as 
\begin{equation}
\frac{1}{2\pi}\mathcal{H}_{m}=\sum_{j}^{n}[(-\nu_{C}-\frac{A_{zz}^{j}}{2})I_{z}^{j}+A_{zz}^{j}s_{z}I_{z}^{j}+A_{zx}^{j}s_{z}I_{x}^{j}-\frac{A_{zx}^{j}}{2}I_{x}^{j}].\label{Hsubmult}
\end{equation}
Here $I_{z/x}^{j}$ denotes the $z/x$ spin operator for $j$th $^{13}$C
spin, and $A_{zz/zx}^{j}$ the couplings with the electron.

 Figure \ref{quancoup} shows the dependence of the $^{13}$C quantization
axis orientation $\theta_{-}$ in the subspace of electron state $m_{S}=-1$
on the coupling strengths. Since for $m_{S}=0$ the quantization axis
is aligned with the $z$-axis, $\theta_{-}$ is also the difference
between the orientations. Here we only consider angles close to $90^{\circ}$,
since these values offer high control efficiency \citep{PhysRevA.76.032326}.
The couplings we used in the simulation are indicated by circles in
Figure \ref{quancoup}, and we list the values in Table \ref{quantmulC}.
\begin{table}
\begin{tabular}{|c|c|c|c|}
\hline 
$^{13}$C number $j$  & $A_{zz}^{j}$ (MHz)  & $A_{zx}^{j}$ (MHz)  & Quantization axis $\theta_{-}$ \tabularnewline
\hline 
$1$  & $-0.152$  & $0.110$  & $87^{\circ}$ \tabularnewline
\hline 
$2$  & $-0.198$  & $0.328$  & $97^{\circ}$\tabularnewline
\hline 
$3$  & $-0.228$  & $0.164$  & $113^{\circ}$\tabularnewline
\hline 
$4$  & $-0.304$  & $0.274$  & $118^{\circ}$ \tabularnewline
\hline 
\end{tabular}\caption{Couplings of the $j^{\text{th }}$ $^{13}$C with the electron spin. }
\label{quantmulC} 
\end{table}

We still use the method presented in Section \ref{subsec:Pulse-Sequences}
to optimize the pulse sequence. Here we use 3-4 MW pulses with 4-5
delays to implement the controlled-$R_{x}^{j}(\pi)$ with one $^{13}$C
spin as the target qubit, where $R_{x}^{j}(\pi)=e^{-i\pi I_{x}^{j}}$
with $j$ indicating the affected $^{13}$C spin. The Rabi frequency
is 0.5 MHz for 2-4 qubit system, but 1 MHz in the 5 qubit system.
The parameters and the obtained fidelities are listed in Table \ref{tabparamulti}.

\begin{table*}
\begin{subtable}{0.45\textwidth} \centering %
\begin{tabular}{|c|c|c|}
\hline 
Delay ($\mu$s)  & Pulse duration ($\mu$s)  & Phase ($^{\circ}$) \tabularnewline
\hline 
$\tau_{0}=3.452$  &  & \tabularnewline
\hline 
$\tau_{1}=2.059$  & $t_{1}=1.910$  & $\phi_{1}=179$ \tabularnewline
\hline 
$\tau_{2}=2.124$  & $t_{2}=3.888$  & $\phi_{2}=136$ \tabularnewline
\hline 
$\tau_{3}=1.000$  & $t_{3}=1.915$  & $\phi_{3}=90$ \tabularnewline
\hline 
\end{tabular}\caption{Controlled-$R_{x}^{1}(\pi)$ in 2 qubits, with fidelity as 0.997. }
\label{tabparamcont} \end{subtable} 
\begin{subtable}{0.45\textwidth} \centering %
\begin{tabular}{|c|c|c|}
\hline 
Delay ($\mu$s)  & Pulse duration ($\mu$s)  & Phase ($^{\circ}$) \tabularnewline
\hline 
$\tau_{0}=3.294$  &  & \tabularnewline
\hline 
$\tau_{1}=1.304$  & $t_{1}=0.766$  & $\phi_{1}=284$ \tabularnewline
\hline 
$\tau_{2}=2.707$  & $t_{2}=0.222$  & $\phi_{2}=235$ \tabularnewline
\hline 
$\tau_{3}=2.952$  & $t_{3}=1.160$  & $\phi_{3}=94$ \tabularnewline
\hline 
$\tau_{4}=2.463$  & $t_{4}=3.006$  & $\phi_{4}=90$ \tabularnewline
\hline 
\end{tabular}\caption{Controlled-$R_{x}^{1}(\pi)$ in 3 qubits, with fidelity as 0.995. }
\label{tabparamcnot3_12} \end{subtable} 

\bigskip{}
\begin{subtable}{0.45\textwidth} \centering %
\begin{tabular}{|c|c|c|}
\hline 
Delay ($\mu$s)  & Pulse duration ($\mu$s)  & Phase ($^{\circ}$) \tabularnewline
\hline 
$\tau_{0}=1.070$  &  & \tabularnewline
\hline 
$\tau_{1}=1.679$  & $t_{1}=3.612$  & $\phi_{1}=87$ \tabularnewline
\hline 
$\tau_{2}=3.071$  & $t_{2}=3.924$  & $\phi_{2}=263$ \tabularnewline
\hline 
$\tau_{3}=3.711$  & $t_{3}=0.370$  & $\phi_{3}=224$ \tabularnewline
\hline 
$\tau_{4}=3.702$  & $t_{4}=0.415$  & $\phi_{4}=90$ \tabularnewline
\hline 
\end{tabular}\caption{Controlled-$R_{x}^{2}(\pi)$ in 3 qubits, with fidelity as 0.995. }
\label{tabparamcnot3_13} \end{subtable} 
\begin{subtable}{0.45\textwidth} \centering %
\begin{tabular}{|c|c|c|}
\hline 
Delay ($\mu$s)  & Pulse duration ($\mu$s)  & Phase ($^{\circ}$) \tabularnewline
\hline 
$\tau_{0}=1.384$  &  & \tabularnewline
\hline 
$\tau_{1}=1.615$  & $t_{1}=2.163$  & $\phi_{1}=113$ \tabularnewline
\hline 
$\tau_{2}=3.286$  & $t_{2}=0.133$  & $\phi_{2}=15$ \tabularnewline
\hline 
$\tau_{3}=5.199$  & $t_{3}=1.126$  & $\phi_{3}=141$ \tabularnewline
\hline 
$\tau_{4}=1.375$  & $t_{4}=1.202$  & $\phi_{4}=90$ \tabularnewline
\hline 
\end{tabular}\caption{Controlled-$R_{x}^{1}(\pi)$ in 4 qubits, with fidelity as 0.997. }
\label{tabparamcnot4_12} \end{subtable} 
\begin{subtable}{0.45\textwidth} \centering %
\begin{tabular}{|c|c|c|}
\hline 
Delay ($\mu$s)  & Pulse duration ($\mu$s)  & Phase ($^{\circ}$) \tabularnewline
\hline 
$\tau_{0}=0.981$  &  & \tabularnewline
\hline 
$\tau_{1}=2.490$  & $t_{1}=0.963$  & $\phi_{1}=253$ \tabularnewline
\hline 
$\tau_{2}=5.768$  & $t_{2}=1.543$  & $\phi_{2}=202$ \tabularnewline
\hline 
$\tau_{3}=1.411$  & $t_{3}=0.370$  & $\phi_{3}=72$ \tabularnewline
\hline 
$\tau_{4}=4.837$  & $t_{4}=0.765$  & $\phi_{4}=90$ \tabularnewline
\hline 
\end{tabular}\caption{Controlled-$R_{x}^{2}(\pi)$ in 4 qubits, with fidelity as 0.991. }
\label{tabparamcnot4_13} \end{subtable} \begin{subtable}{0.45\textwidth}
\centering %
\begin{tabular}{|c|c|c|}
\hline 
Delay ($\mu$s)  & Pulse duration ($\mu$s)  & Phase ($^{\circ}$) \tabularnewline
\hline 
$\tau_{0}=1.277$  &  & \tabularnewline
\hline 
$\tau_{1}=1.742$  & $t_{1}=0.758$  & $\phi_{1}=212$ \tabularnewline
\hline 
$\tau_{2}=2.903$  & $t_{2}=0.751$  & $\phi_{2}=60$ \tabularnewline
\hline 
$\tau_{3}=0.744$  & $t_{3}=1.076$  & $\phi_{3}=87$ \tabularnewline
\hline 
$\tau_{4}=0.719$  & $t_{4}=1.490$  & $\phi_{4}=90$ \tabularnewline
\hline 
\end{tabular}\caption{Controlled-$R_{x}^{3}(\pi)$ in 4 qubits, with fidelity as 0.956. }
\label{tabparamcnot4_14} \end{subtable} 
\begin{subtable}{0.45\textwidth} \centering %
\begin{tabular}{|c|c|c|}
\hline 
Delay ($\mu$s)  & Pulse duration ($\mu$s)  & Phase ($^{\circ}$) \tabularnewline
\hline 
$\tau_{0}=3.428$  &  & \tabularnewline
\hline 
$\tau_{1}=4.375$  & $t_{1}=0.354$  & $\phi_{1}=226$ \tabularnewline
\hline 
$\tau_{2}=0.665$  & $t_{2}=1.731$  & $\phi_{2}=169$ \tabularnewline
\hline 
$\tau_{3}=1.707$  & $t_{3}=0.631$  & $\phi_{3}=205$ \tabularnewline
\hline 
$\tau_{4}=2.060$  & $t_{4}=1.674$  & $\phi_{4}=90$ \tabularnewline
\hline 
\end{tabular}\caption{Controlled-$R_{x}^{1}(\pi)$ in 5 qubits, with fidelity as 0.996. }
\label{tabparamcnot5_12} \end{subtable} 
\begin{subtable}{0.45\textwidth} \centering %
\begin{tabular}{|c|c|c|}
\hline 
Delay ($\mu$s)  & Pulse duration ($\mu$s)  & Phase ($^{\circ}$) \tabularnewline
\hline 
$\tau_{0}=3.290$  &  & \tabularnewline
\hline 
$\tau_{1}=0.835$  & $t_{1}=1.865$  & $\phi_{1}=77$ \tabularnewline
\hline 
$\tau_{2}=5.360$  & $t_{2}=1.077$  & $\phi_{2}=138$ \tabularnewline
\hline 
$\tau_{3}=0.814$  & $t_{3}=0.572$  & $\phi_{3}=97$ \tabularnewline
\hline 
$\tau_{4}=4.277$  & $t_{4}=2.477$  & $\phi_{4}=90$ \tabularnewline
\hline 
\end{tabular}\caption{Controlled-$R_{x}^{2}(\pi)$ in 5 qubits, with fidelity as 0.987. }
\label{tabparamcnot5_13} \end{subtable} \begin{subtable}{0.45\textwidth}
\centering %
\begin{tabular}{|c|c|c|}
\hline 
Delay ($\mu$s)  & Pulse duration ($\mu$s)  & Phase ($^{\circ}$) \tabularnewline
\hline 
$\tau_{0}=1.903$  &  & \tabularnewline
\hline 
$\tau_{1}=3.557$  & $t_{1}=2.005$  & $\phi_{1}=84$ \tabularnewline
\hline 
$\tau_{2}=3.055$  & $t_{2}=1.248$  & $\phi_{2}=323$ \tabularnewline
\hline 
$\tau_{3}=2.940$  & $t_{3}=1.821$  & $\phi_{3}=334$ \tabularnewline
\hline 
$\tau_{4}=3.984$  & $t_{4}=2.386$  & $\phi_{4}=90$ \tabularnewline
\hline 
\end{tabular}\caption{Controlled-$R_{x}^{3}(\pi)$ in 5 qubits, with fidelity as 0.942. }
\label{tabparamcnot5_14} \end{subtable} \begin{subtable}{0.45\textwidth}
\centering %
\begin{tabular}{|c|c|c|}
\hline 
Delay ($\mu$s)  & Pulse duration ($\mu$s)  & Phase ($^{\circ}$) \tabularnewline
\hline 
$\tau_{0}=0.112$  &  & \tabularnewline
\hline 
$\tau_{1}=1.927$  & $t_{1}=1.417$  & $\phi_{1}=100$ \tabularnewline
\hline 
$\tau_{2}=2.568$  & $t_{2}=2.062$  & $\phi_{2}=30$ \tabularnewline
\hline 
$\tau_{3}=1.562$  & $t_{3}=0.555$  & $\phi_{3}=10$ \tabularnewline
\hline 
$\tau_{4}=1.377$  & $t_{4}=1.895$  & $\phi_{4}=90$ \tabularnewline
\hline 
\end{tabular}\caption{Controlled-$R_{x}^{4}(\pi)$ in 5 qubits, with fidelity as 0.930. }
\label{tabparamcnot5_15} \end{subtable}

\caption{Parameters of the sequences to implement controlled-$R_{x}^{j}(\pi)$
in the hybrid system consisting of 2-5 qubits.}
\label{tabparamulti} 
\end{table*}

\end{enumerate}

\end{document}